\documentclass[aps,prx,reprint,superscriptaddress]{revtex4-2}

\usepackage{amsmath,amssymb,siunitx,physics,graphicx}
\graphicspath{{./images/}}

\begin{document}


\title{The role of Rashba spin-orbit induced spin textures in the anomalous Josephson effect.}

\author{R. D. Monaghan}
\email[]{ross.monaghan@adelaide.edu.au}
\affiliation{School of Physical Sciences, The University of Adelaide}
\affiliation{School of Chemical and Advanced Engineering, The University of Adelaide}

\author{G. C. Tettamanzi}
\affiliation{School of Physical Sciences, The University of Adelaide}
\affiliation{School of Chemical and Advanced Engineering, The University of Adelaide}

\date{\today}

\begin{abstract}
This work reports the theoretical investigation into the mechanism underpinning the anomalous Josephson effect. The prototypical system we study is a ballistic two-dimensional junction containing a two-dimensional Rashba spin-orbit interaction. In this paper we demonstrate how this two-dimensional Rashba interaction mixes the spins of adjacent transverse subbands which leads to significant spin-asymmetry within the junction. Under an external magnetic field, applied perpendicular to both the axis of transport and the normal vector of the junction, the sinusoidal Josephson current can then experience an anomalous phase shift. The role of this spin mixing in the limit of a single sub-band is initially explored by deriving an analytical expression for the resulting anomalous phase shift. The analysis is then extended to systems with multiple occupied sub-bands; in this later section, starting from a microscopic model, we derive an analytic formula for the resulting anomalous phase shift indicating it is linear in both magnetic field and spin-orbit strength. We then verify and validate all findings by comparing them with numerical results evaluated by a tight-binding model.

\end{abstract}

\maketitle

\section{Introduction}\label{sec::Intro}

    The Josephson effect is one of the most profound results concerning the phase of quantum objects; for a phase difference of $\varphi$ between two coupled superconductors, separated by a region of non-superconducting material, a sinusoidal electrical current will flow \cite{josephsonPossibleNewEffects1962}. A long-standing result in the field of Josephson junctions was that applying an appropriately large external magnetic field to the junction could induce a $\pi$ phase shift in the sinusoidal current, where the minimum of the free energy of the system now occurs when $\varphi = \pi$ \cite{bulaevskiiSuperconductingSystemWeak1977}. A natural extension of this concept is to construct a junction where the Josephson current acquires an anomalous phase factor $\varphi_0$, with $0 \leq \varphi_0 \leq 2\pi$, such that the total current as a function of the phase difference $\varphi$ across the superconductors is given by
        \begin{equation}\label{eq::Anom}
            I = I_c \sin(\varphi + \varphi_0)  \; .
        \end{equation} 
    To generate such an arbitrary phase shift it was realised that, regardless of the physical mechanism, both time-reversal and spatial-inversion symmetry of the system must be broken \cite{rasmussenEffectsSpinorbitCoupling2016}. One way to do so is by constructing Josephson junctions where the non-superconducting section of the junction is made from material possessing a strong Rashba spin-orbit signature, and by also applying an external magnetic field perpendicular to both the axis of transport and the normal of the substrate. Careful experimental work over the recent years has indeed detected this anomalous phase in systems possessing both Rashba spin-orbit coupling and an external magnetic field \cite{assoulineSpinOrbitInducedPhaseshift2019,strambiniJosephsonPhaseBattery2020,reinhardtLinkSupercurrentDiode2023,mayerGateControlledAnomalous2020}. This effect is now known as the anomalous Josephson effect. However, no clear and unified explanation has been proposed regarding the microscopic nature of this effect \cite{shukrinovAnomalousJosephsonEffect2022}.

    In quasi one-dimensional devices, such as the ones studied in this work, the physical confinement can give rise to multiple transverse subbands. A well known result is that under a two-dimensional Rashba spin-orbit interaction, these transverse subbands can be coupled and the spins of the states mixed \cite{morozEffectSpinorbitInteraction1999,berciouxQuantumTransportRashba2015}. The resulting mixing of the spin states -- which we denote as the spin-texture of the system -- has been theoretically investigated previously within the context of non-superconducting transport \cite{governaleSpinAccumulationQuantum2002,muraniAndreevSpectrumHigh2017,parkAndreevSpinQubits2017,reynosoAnomalousJosephsonCurrent2008,reynosoSpinorbitinducedChiralityAndreev2012}. This mixing has also recently been observed in experimental spectroscopy measurements on  two-dimensional Josephson junction \cite{tosiSpinOrbitSplittingAndreev2019}. The purpose of this work is to explicitly investigate the effects of the spin-texture on the resulting anomalous phase -- deriving expressions for the anomalous phase as a function of the spin-texture and, finally, to provide a more complete microscopic model that can justify the appearance of this effect in different systems. Although previous work has considered the role of subband mixing within the framework of the anomalous Josephson effect (see Refs \cite{kriveChiralSymmetryBreaking2004, reynosoAnomalousJosephsonCurrent2008,reynosoSpinorbitinducedChiralityAndreev2012,yokoyamaAnomalousJosephsonEffect2014}), these papers either only consider the Fermi velocity asymmetry induced by the subband mixing, or as in the case of Ref \cite{reynosoAnomalousJosephsonCurrent2008}, discuss how the mixing of the spins can act as a spin-polariser to generate the anomalous phase. 
    
    In this work, we initially study single-channel systems with two-dimensional Rashba interactions such that we can generate analytic expressions to clearly demonstrate the physics. The results are then extended to multi-channel systems where it is possible, after some approximations, to make experimental predictions; to verify these predictions, the anomalous phase shift is then explicitly computed numerically through the use of the non-equilibrium Green function formalism. 

\section{Model for the anomalous Josephson effect}\label{sec::AnomalousJosephsonModel}

    Within the literature on the anomalous Josephson effect there are currently two leading microscopic mechanisms which can be used to explain the observed experimental results. The first is spin-related interference effects induced by the combination of Zeeman and Rashba spin-orbit coupling which were first introduced by Krive \textit{et al.} in Ref. \cite{kriveChiralSymmetryBreaking2004}. However, another mechanism proposed is orbital-related interference effects induced by the vector potential and disorder within the system as in Refs \cite{mironovDoublePathInterference2015, zuoSupercurrentInterferenceFewMode2017}. Indeed, based purely on symmetry arguments, Rasmussen \textit{et al.} showed that either spin-interference, or orbital-interference, could induce a non-zero supercurrent at zero phase difference \cite{rasmussenEffectsSpinorbitCoupling2016}. In this work, our focus will be purely on the role of spin-channel interference and, as a result, both disorder and vector potential will be ignored within our model. Strictly speaking, this restricts our analysis to systems where the total magnetic flux piercing the non-superconducting region is much smaller than a flux quantum $\Phi_0$ such that any phase gained by the vector potential is negligible. Similarly, to ignore disorder we only consider systems which are purely in the ballistic limit.

    \subsection{Single particle Hamiltonian}\label{ssec::SingleParticleHam}    

        \begin{figure}[ht]
            \centering
            \includegraphics[width=0.5\textwidth]{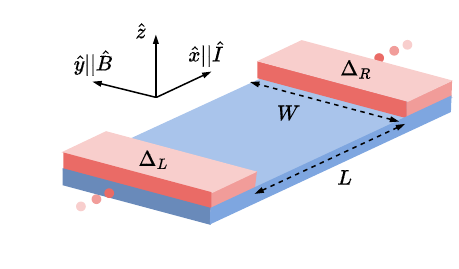}
            \caption{Diagram of the prototypical system studied throughout this paper. It consists of a two-dimensional material which, due to the confinement $W$ in the $y$-axis is quasi one-dimensional along the $x$-axis. The blue material is some semiconductor which has a Rashba spin-orbit interaction present, and is assumed to extend to infinity in the $x$-axis. The red regions are some Type-I superconductor which has been deposited onto the semiconducting layer such that the left and right regions of the device contain some proximitised superconductivity. The uncovered central region of the Josephson junction has length $L$.}
            \label{fig::roughDiagram}
        \end{figure} 

    We consider the prototypical Josephson junction with a ballistic non-superconducting central region sandwiched between two superconducting regions depicted in Fig. \ref{fig::roughDiagram}. The blue region is some non-superconducting semi-conducting material which possesses a two-dimensional Rashba spin-orbit interaction. The red regions in Fig. \ref{fig::roughDiagram} illustrate some Type-I superconducting material which has been deposited over the semiconductor to induce semi-infinite superconducting leads by the proximity effect \cite{degennesBoundaryEffectsSuperconductors1964}. Conceptually, in our model the system consists of three regions: the `left' and `right' proximitised regions are those covered by a superconducting layer and are assumed to extend to infinity, whilst the non-proximitised uncovered region is known as the `central' region. The entire device is assumed to be quasi two-dimensional with a hard-wall boundary conditions confining the system in the $y$-axis. We also assume that the superconducting coherence length $\xi$ is larger than the length $L$ of the central region such that our system is in the short junction limit \cite{beenakkerThreeUniversalMesoscopic1992}. 
    
    We assume that an external magnetic field is applied along the $y$ axis such that it is perpendicular to both the normal of the substrate and the direction of transport -- this is depicted in Fig. \ref{fig::roughDiagram}. We also assume that the magnetic flux is entirely screened from the left and right regions due to the deposited superconducting layer. As a result, the single particle Hamiltonian in the central non-proximitised region can be written as  
        \begin{equation}\label{eq::CentralHam}
            H_C = H_0 + H_{\alpha} + H_Z  \; ,
        \end{equation} 
    where the free single-particle Hamiltonian is given by
        \begin{equation}\label{eq::bareHam}
            H_0 = -\frac{\hbar^{2}\nabla^{2}}{2m^*} - \mu  \; ,
        \end{equation} 
    and $m^*$ denotes the effective mass for the semiconductor being modelled, whilst $\mu$ is the Fermi level. The two-dimensional Rashba spin-orbit term is written as \cite{morozEffectSpinorbitInteraction1999}
        \begin{equation}\label{eq::SingleParticleRashba}
            H_{\alpha} = i\alpha\sigma_y \partial_x - i\alpha \sigma_x \partial_y  \; ,
        \end{equation} 
    where $\alpha$ is the Rashba spin-orbit strength, and we have used the usual Pauli matrices to span spin-space. The Zeeman term is written as
        \begin{equation}\label{eq::SingleParticleZeeman}
            H_Z = E_Z \sigma_y  \; ,
        \end{equation} 
    where $E_Z = g\mu_B B / 2$ for some material dependent $g$ factor. We have ignored any on-site scattering potentials, placing ourselves purely in the ballistic limit. Furthermore, as our system is two-dimensional the height $H$ of the device in the $z$-axis is small such that the flux penetrating is negligible, $BLH \ll h / e$. As a result, it is reasonable to ignore the vector potential \cite{vanheckZeemanSpinorbitEffects2017}.

    For the left and right regions of the device, the non-superconducting component of the Hamiltonian is given by
        \begin{equation}\label{eq::LeftRightElectronic}
            H_{L/R} = H_0 + H_{\alpha}  \; ,
        \end{equation} 
    where we have omitted the Zeeman term as we assume that, due to the deposited superconducting layer, the magnetic field has been entirely expelled by the Meissner effect. To account for the superconducting proximity effect induced by the deposited superconductors, we introduce a superconducting order parameter $\Delta$ which acts microscopically to couple the electrons and holes of the system \cite{rammerQuantumFieldTheory2007}. As a result, we enlarge our basis such that the order parameter couples the non-superconducting Hamiltonian with its time reversed pair. Considering this, the Hamiltonian in the proximitised semi-conducting leads takes the form
        \begin{equation}\label{eq::LeftRightFull}
            H_{L/R} \to \begin{pmatrix}
                H_0 + H_{\alpha}   & i\sigma_y \Delta_{L/R}\\
                -i\sigma_y \Delta_{L/R} & -(H_0 + H_{\alpha})^*
            \end{pmatrix}  \; .
        \end{equation} 
    \subsection{Spin-textures in wavevector space}\label{ssec::SpinTextures}

\begin{figure}[ht]
            \centering
            \includegraphics[width=0.5\textwidth]{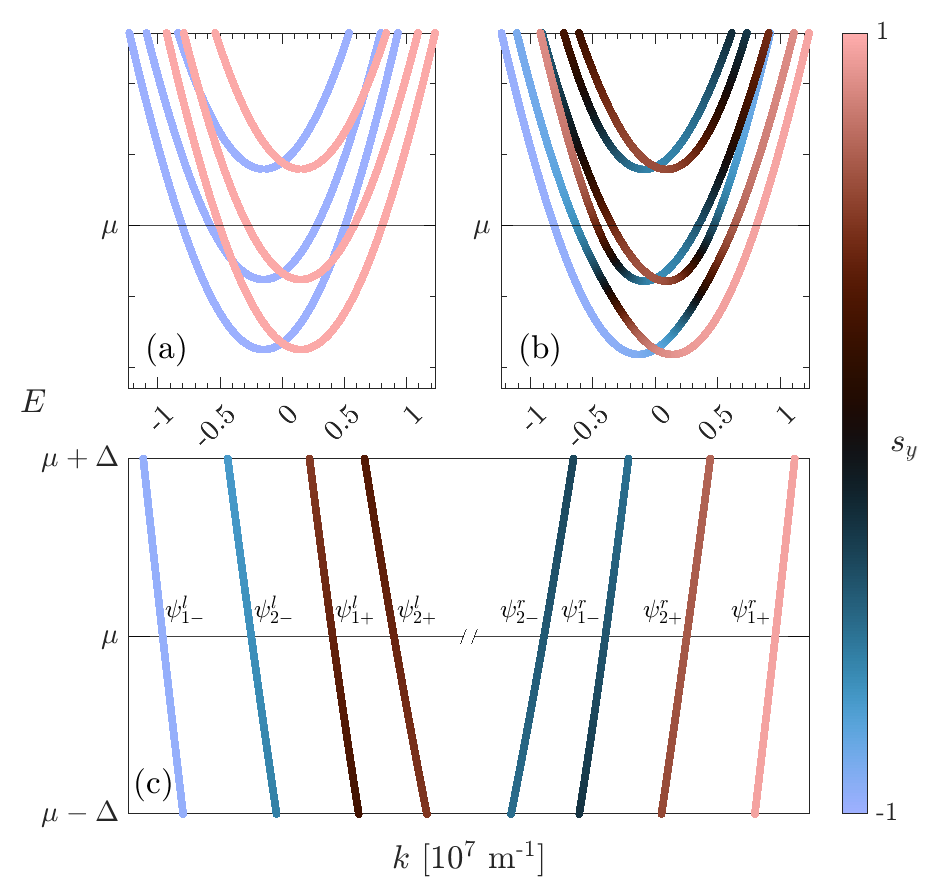}
            \caption{ Numerical dispersion relation computed used the tight-binding techniques of Ref. \cite{mirelesBallisticSpinpolarizedTransport2001}. The wavevector $k$ denotes the wavevector along the $x$-axis. The colours of the curves indicate the corresponding numerical eigenstate projected onto the $y$-axis, $\left< \sigma_y\right>$. No magnetic field is applied to this system.  (a) The dispersion relation for a quasi one-dimensional system with a purely one-dimensional Rashba spin-orbit coupling. (b) The dispersion relation for a quasi one-dimensional system with a physically correct two-dimensional Rashba spin-orbit coupling. (c) The linear states labelled for the energies around the Fermi level.}
            \label{fig::RashbaSpectrum}
        \end{figure}

        In a purely one-dimensional system, the Rashba spin-orbit term of Eq. \ref{eq::SingleParticleRashba} can be simplified to the form $H_{\alpha}^{1\text{D}} = i\alpha\sigma_y \partial_x$ \cite{pedderDynamicResponseFunctions2016}. Within this one-dimensional system, the full Hamiltonian commutes with the $y$ component of the electron's spin $\left[H_0 +H_\alpha + H_Z,\sigma_y\right] = 0$. If we now include a transverse dimension, such that the system has a non-zero length $L$ and width $W$, the confinement in the $y$-axis will produce subbands which can be labelled by an index characterising the discrete states within an infinite potential well. Retaining the purely one-dimensional Rashba spin-orbit term, the Hamiltonian is entirely separable in $x$ and $y$ such that both the spin and subband indices are good quantum numbers. The resulting dispersion relation for this system with zero external magnetic field is illustrated in Fig. \ref{fig::RashbaSpectrum}.a. This has been computed numerically with a tight-binding calculation described in Ref. \cite{mirelesBallisticSpinpolarizedTransport2001}. The colours depict the value of the corresponding numerical eigenstate projected onto the $y$-axis, $\left< \sigma_y\right>$, which we denote as the spin-projection \cite{berciouxQuantumTransportRashba2015}. As this is essentially the dispersion relation of an electron waveguide with only transport in the $x$-axis, we refer to the wavevector along the $x$-axis as simply $k$, rather than $k_x$.
        %

        Importantly, a physically two-dimensional system with a one-dimensional Rashba spin-orbit interaction is not realistic; to accurately model this system we need to write the full two-dimensional Rashba spin-orbit interaction as
            \begin{equation}\label{eq::TwoDRashbaEx}
                H^{2\text{D}}_\alpha =  i\alpha\sigma_y \partial_x - i\alpha \sigma_x \partial_y  \; .
            \end{equation} 
        The immediate result of the extra term containing a $\sigma_x$ matrix, is that $\sigma_y$ no longer commutes with the Hamiltonian. This well-known result leads to the mixing of states of opposite spin and adjacent subband index \cite{morozEffectSpinorbitInteraction1999}. The dispersion relation for this system is sketched in Fig. \ref{fig::RashbaSpectrum}.b. As spin is no longer a good quantum number, where there was previously a crossing between odd and even numbered subbands of opposite spin there is now an avoided crossing where the spin-projection smoothly flips. 

        To clarify the eventual role of this spin-texture, it is useful to linearise the dispersion relation around the Fermi wavevectors for each state -- the Fermi wavevector for the $j$\textsuperscript{th} state being denoted by $k_j$. As the transport in the quasi one-dimensional structure is entirely along the $x$-axis, we can decompose the electronic eigenstates $\ket{\psi_{j\pm}}$ into a linear combination of left-moving, $\ket{\psi^l_{j\pm}}$, and right-moving, $\ket{\psi^r_{j\pm}}$, components \cite{vanheckZeemanSpinorbitEffects2017}
        \begin{equation}\label{eq::leftMoving}
            \ket{\psi_{j\pm}} = e^{-ik_j x}\ket{\psi^l_{j\pm}} + e^{ik_j x}\ket{\psi^r_{j\pm}}  \; .
        \end{equation} 
    %
        Although the subband index $j$, and spin index $\pm$ are no longer good quantum numbers, to clarify the notation we can still denote each state by the indices it would have in the no Rashba limit $\alpha\to 0$. We sketch the linearised dispersion relation in Fig. \ref{fig::RashbaSpectrum}.c where we have also labelled each state by their three indices: subband, spin, and direction. The linearised dispersion relation for each state is entirely characterised by three parameters -- namely, the Fermi wavevector of each state $k_{j\pm}^{l/r}$ defined such that
            \begin{equation}\label{eq::defFermi}
                E(k_{j\pm}^{l/r})- \mu = 0  \; ,
            \end{equation} 
        the Fermi velocity, which is proportional to the slope of the linearised spectrums
            \begin{equation}\label{eq::FermiVelocityDefinition}
                v_{j\pm}^{l/r} \equiv \frac{1}{\hbar} \left. \frac{\partial E}{\partial k}\right|_{k =  k_{j;\pm}^{l/r}}  \; ,
            \end{equation} 
        and the spin-projection which is given by the overlap
        \begin{equation}\label{eq::LinearisedSpinTextureDefinition}
            s_{j\pm}^{l/r} \equiv \expval{\sigma_y}{\psi_{j\pm}^{l/r}}  \; .
        \end{equation} 
        %
        Importantly, in the absence of an external magnetic field, the system still possesses time-reversal symmetry such that \cite{parkAndreevSpinQubits2017}
            \begin{equation}\label{eq::TimeReversalSymmetry}
                    \begin{aligned}
                        E(k) &= E(-k)  \; ,\\
                        \ket{\psi^l_{j\pm}} &= \mathcal{T} \ket{\psi^r_{j\mp}} \; ,
                    \end{aligned} 
            \end{equation} 
        where $\mathcal{T}$ is the time reversal operator. As a result of this, a pair of useful identities which are valid in the absence of a magnetic field are that
            \begin{equation}\label{eq::UsefulIdentityTime}
                v^l_{\pm} = -v^r_{\mp} \qquad s^l_{\pm} = -s^r_{\mp}   \; ,
            \end{equation} 
        which can be observed visually in Fig. \ref{fig::RashbaSpectrum}.c.

        To now briefly study how the dispersion relation changes under an external magnetic field applied along the spatial $y$-axis, we restrict ourselves to systems where the magnetic field is small relative to the subband spacing. This is not entirely restrictive, as for say a generic InAs device with a width of $200$nm, this corresponds to magnetic fields less than $400$mT, which is certainly within the range of magnetic fields commonly applied to these systems \cite{strambiniJosephsonPhaseBattery2020,mayerGateControlledAnomalous2020,assoulineSpinOrbitInducedPhaseshift2019,reinhardtLinkSupercurrentDiode2023}. For these small magnetic fields, it is reasonable to assume that the spin-texture and Fermi velocities are unchanged as they only vary appreciably for energies on the scale of the subband spacing \cite{parkAndreevSpinQubits2017}. Furthermore, as the Zeeman term only contains a $\sigma_y$ matrix, then the Zeeman term expressed in the basis of the linearised states $\ket{\psi^{l/r}_{j\pm}}$ is given simply by $H_Z = s^{l/r}_{j\pm}E_Z$, where we have substituted in the definition of spin-projection defined in Eq. \ref{eq::LinearisedSpinTextureDefinition}. As a result, we can express the linearised energy spectrums as 
            \begin{equation}\label{eq::LinearisedEquations}
                \varepsilon_{j\pm}^{l/r}(k) =  \hbar v_{j\pm}^{l/r}k - s_{j\pm}^{l/r}E_Z \; .
            \end{equation} 
    \subsection{Anomalous phase induced by Andreev bound states}\label{ssec::ABS}

\begin{figure}[ht]  
                \centering
                \includegraphics[width=0.5\textwidth]{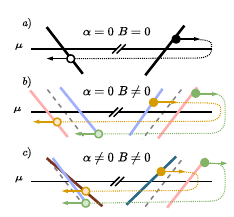}
                \caption{(a) The diagram of an Andreev reflection, where a right-moving electron is Andreev reflected as a hole below the Fermi level. With no terms acting on spin, the linearised dispersion relation is spin degenerate. (b) Under a magnetic field, and with no spin-orbit coupling, the dispersion relation is Zeeman split such that two Andreev bound states appear. Although the resulting Andreev bound states are phase shifted, due to the opposite spins the two phases cancel each other out so that the total system gains no anomalous phase -- illustrated by both Andreev bound states having the same path length. (c) Under both a two-dimensional spin-orbit term and a magnetic field, the asymmetry in the spin-projections will induce an anomalous phase -- illustrated by the different path lengths. The colours of the states correspond to the magnitude of the spin-projection; for reference see the colourbar in Fig. \ref{fig::RashbaSpectrum}.}
                \label{fig::ABSdiagram}
            \end{figure}

        As introduced in Eq. \ref{eq::LeftRightFull}, the order parameter acts to couple the electrons and holes within the superconducting system. As a result of this, an electron in the non-proximitised region of the Josephson junction which is incident on the superconducting interface will be Andreev reflected as its time reversed partner \cite{andreevThermalConductivityIntermediate1964,saulsAndreevBoundStates2018}. For a Josephson junction, where the non-proximitised region is sandwiched between two superconductors, this leads to the formation of bound states consisting of counter-propagating electrons and holes -- an object known as an Andreev bound state \cite{kulikMacroscopicQuantizationProximity1969,bardeenJosephsonCurrentFlow1972}. This process is illustrated in Fig. \ref{fig::ABSdiagram}.a. It is important to remember that in this work we focus solely on the short junction limit where the superconducting coherence length $\xi$ is much greater than the length $L$ of the junction; in this limit the supercurrent is entirely carried by the Andreev bound states rather than states outside of the superconducting gap \cite{beenakkerUniversalLimitCriticalcurrent1991}. 

        The Andreev bound states which carry the Josephson current are entirely formed by the states within the non-proximitised central region. Hence, having discussed the microscopic spin-texture of the states within the non-proximitised central region in Sec. \ref{ssec::SpinTextures}, we can now discuss the ramifications of the spin-texture on the resulting Andreev bound states. To begin, previous literature has shown that, generically, to observe the anomalous Josephson effect you need to introduce some asymmetry between different Andreev bound states of the system \cite{kriveChiralSymmetryBreaking2004,nesterovAnomalousJosephsonEffect2016}. For example, if we consider the case of no Rashba spin-orbit coupling, but a non-zero external magnetic field -- illustrated in Fig. \ref{fig::ABSdiagram}.b -- although we see the spin-bands shift, no asymmetry is generated. As a result, we expect no anomalous effect. However, if we now include the two-dimensional Rashba spin-orbit coupling, in conjunction with the magnetic field along the $y$-axis -- illustrated in Fig. \ref{fig::ABSdiagram}.c -- the resulting Fermi velocity asymmetry and spin-texture, mixed with the external magnetic field, induces significant asymmetry in the resulting pair of Andreev bound states. Although previous theoretical work has studied this asymmetry, they ignored the role of the spin-texture focusing instead only on the Fermi velocities \cite{kriveChiralSymmetryBreaking2004, yokoyamaAnomalousJosephsonEffect2014}. The main focus of this work is to provide both a quantitative and qualitative link between the spin-texture asymmetry induced by the two-dimensional Rashba spin-orbit coupling and the anomalous Josephson effect.
        

        To understand the consequences of this asymmetry in our system, we now look to evaluate the energy of the different Andreev bound states. A simplification of working in the short junction limit is that each linearised state $\psi^{l/r}_{j\pm}$ will give rise to only a single Andreev bound state \cite{saulsAndreevBoundStates2018}. Indeed, an Andreev bound state with energy $\varepsilon$ can only be formed by matching the wavefunctions of both the linearised state $\psi^{l/r}_{j\pm}$ with energy $\varepsilon$ above the Fermi level, to both its time-reversed hole partner, and to the exponentially decaying wavefunctions within the superconducting leads \cite{saulsAndreevBoundStates2018}. The details are left for Appendix. \ref{sec::ABSformation}, however, doing so leads to the following transcendental equation which the energy $\varepsilon$ of the resulting Andreev bound state must satisfy
            \begin{equation}\label{eq::TransverseEq}
                e^{-i\varphi}e^{-2i\text{arccos}(\varepsilon / |\Delta|)} e^{2iL\varepsilon / \hbar v_{j\pm}^{l/r}}e^{2iLs_{j\pm}^{l/r}E_{Z} / \hbar v_{j\pm}^{l/r}}  = 1 \; .
            \end{equation} 
        Taking the log of both sides we find a quantisation condition for the phase acquired by the counter-propagating electrons and holes forming the Andreev bound state
        \begin{equation}\label{eq::PhaseQuantisationCOndition}
            \varphi + 2\text{arccos}\left(\frac{\varepsilon}{ |\Delta|}\right) - \frac{2L\varepsilon}{\hbar v_{j\pm}^{l/r}} - \frac{2Ls_{j\pm}^{l/r}E_{Z}}{\hbar v_{j\pm}^{l/r}}  = 2\pi n \; ,
        \end{equation} 
        for some integer $n$. The term $\varphi$ is the superconducting phase difference picked up during the Andreev reflection process, the $2\text{arccos}(\varepsilon / |\Delta|)$ term is the phase picked up from the evanescent states in the superconductor, whilst $2L\varepsilon / \hbar v_{j\pm}^{l/r}$ is the phase picked up by traversing the junction of length $L$ \cite{bagwellSuppressionJosephsonCurrent1992}. The interesting term is ${2Ls_{j\pm}^{l/r}E_{Z}} / {\hbar v_{j\pm}^{l/r}}$ which is the phase picked up by the spin moving through the magnetic field. 
        
        As we are working in the short junction limit such that $\xi_{j\pm}^{l/r} \gg L$, where $\xi_{j\pm}^{l/r} \equiv \hbar v_{j\pm}^{l/r} / \Delta$ is the coherence length of that subband state, then the phase picked up by traversing the junction is negligible
            \begin{equation}\label{eq::ShortJunction}
                \frac{L\varepsilon}{\hbar v_{j\pm}^{l/r}} =    \frac{L}{\xi_{j\pm}^{l/r}}\frac{\varepsilon}{|\Delta|} \approx 0 \; .
            \end{equation} 
        As a result, solving for the allowed energies $\varepsilon$ in Eq. \ref{eq::PhaseQuantisationCOndition}, we find that the energy phase relation for the Andreev bound state produced via the $\psi^{l/r}_{j\pm}$ subband state is given by
            \begin{equation}\label{eq::FinalAndreevEnergy} 
                \begin{aligned}
                    \varepsilon^{r}_{j\pm}(\varphi) &= - |\Delta| \cos(\frac{\varphi}{2} - E_ZL\frac{s_{j\pm}^{r}}{\hbar v_{j\pm}^{r}}  )    \; ,\\
                    \varepsilon^{l}_{j\pm}(\varphi) &=  |\Delta| \cos(\frac{\varphi}{2} - E_ZL\frac{s_{j\pm}^{l}}{\hbar v_{j\pm}^{l}}  )  \; .
                \end{aligned} 
            \end{equation} 

        From these bound state energies, we can now directly compute the Josephson current; in the short junction limit the current is entirely given by the contribution through each discrete energy level \cite{beenakkerUniversalLimitCriticalcurrent1991}
            \begin{equation}\label{eq::TotalCurrentEnergy}
                I = -\frac{2e}{\hbar}\sum_{j,l/r,\pm} \left[ \tanh(\frac{\varepsilon_{j\pm}^{l/r}}{2k_B T}) \partialderivative{\varepsilon_{j\pm}^{l/r}}{\varphi} \right]  \; ,
            \end{equation} 
        where $k_BT$ is the thermal energy of the system. Utilising the time-reversal symmetry discussed in Eq. \ref{eq::TimeReversalSymmetry} then we can show that
            \begin{equation}\label{eq::SymmetryInEnergy}
                \varepsilon_{j\pm}^r = -\varepsilon_{j\mp}^l  \; ,
            \end{equation} 
        such that Eq. \ref{eq::TotalCurrentEnergy} can be rewritten to include only right-moving states as
         \begin{equation}\label{eq::TotalCurrentEnergyNice}
            \begin{aligned}
                I = \frac{e|\Delta|^2}{\hbar k_B T}\sum_{j} \left[ \cos(\frac{\varphi}{2} - E_ZL\frac{s_{j+}^{r}}{\hbar v_{j+}^{r}}  )    \sin(\frac{\varphi}{2} - E_ZL\frac{s_{j+}^{r}}{\hbar v_{j+}^{r}}  )   \right] \\  
            + \frac{e|\Delta|^2}{\hbar k_B T}\sum_{j} \left[  \cos(\frac{\varphi}{2} - E_ZL\frac{s_{j-}^{r}}{\hbar v_{j-}^{r}}  ) \sin(\frac{\varphi}{2} - E_ZL\frac{s_{j-}^{r}}{\hbar v_{j-}^{r}}  )  \right] 
            \end{aligned} 
        \end{equation} 
        where we have assumed that we are working near the critical temperature ($k_BT \sim |\Delta|$) to simplify the analytical expression. This is not an essential approximation; however, it allows us to recover a sinusoidal Josephson current even in the ballistic regime \cite{beenakkerThreeUniversalMesoscopic1992}. Using a double angle trigonometric identity this is simplified further to
        \begin{equation}\label{eq::TotalCurrentEnergyNiceTrigIdentities}
            \begin{aligned}
                I = \frac{e|\Delta|^2}{2\hbar k_B T}\sum_{j} \left[    \sin(\varphi- 2E_ZL\frac{s_{j+}^{r}}{\hbar v_{j+}^{r}}  ) \right. \\
                + \left. \sin(\varphi- 2E_ZL\frac{s_{j-}^{r}}{\hbar v_{j-}^{r}}  )    \right] \;.
            \end{aligned} 
        \end{equation} 

        Writing the current as a sum of sinusoids, each with its own phase offset, as in Eq. \ref{eq::TotalCurrentEnergyNiceTrigIdentities}, allows us to explicitly evaluate the anomalous phase $\varphi_0$ by phasor addition. The resulting current can be expressed as
            \begin{equation}\label{eq::TotalCurrentEnergyShift}
                I = I_C\sin(\varphi + \varphi_0)  \; ,
            \end{equation} 
        where $I_C$ is the critical current and the anomalous phase is given by
        \begin{equation}\label{eq::AnomalousPhase} 
            \varphi_0  = \atan\left[\frac{\sum_{j} \left( \sin\frac{ 2E_ZL s_{j+}^{r}}{\hbar v_{j+}^{r}}  +  \sin \frac{ 2E_ZL s_{j-}^{r}}{\hbar v_{j-}^{r}} \right)}{\sum_{j} \left( \cos \frac{ 2E_ZL s_{j+}^{r}}{\hbar v_{j+}^{r}}   +  \cos \frac{ 2E_ZL s_{j-}^{r}}{\hbar v_{j-}^{r}}  \right)} \right] \; .
        \end{equation} 
        We can identify the phase factor ${ 2E_ZL s_{j\pm}^{r}} / {\hbar v_{j\pm}^{r}}$ in this expression from the quantisation condition of Eq. \ref{eq::PhaseQuantisationCOndition} as the phase gained by right-moving electrons with spin-index $\pm$ as they pass through the magnetic field. 
        As we can observe from this expression, the presence of an anomalous phase is solely due to asymmetry in the Fermi velocities and the spin-projections -- namely, there will be an anomalous phase when $v^r_{j+} \neq v^r_{j-}$ or when  $s^r_{j+} \neq -s^r_{j-}$.

\section{Anomalous phase in single-channel systems}\label{sec::AnomalousPhaseSingle}

    In this section the anomalous phase shift is evaluated for systems with only a single occupied subband. The benefit of this system is due to only a pair of spin states being occupied, the resulting physics is simplified and can generally be carried out analytically. We consider a Josephson junction with length $L$ and width $W$ as depicted in Fig. \ref{fig::roughDiagram}. We utilise a two-dimensional Rashba spin-orbit term as given by Eq. \ref{eq::TwoDRashbaEx}.

    \subsection{Effective 1D Hamiltonian}\label{ssec::EffectiveHamiltonian}        
    
         We begin with the same Hamiltonian as outlined in Sec \ref{ssec::SingleParticleHam}, however, we assume the Fermi level $\mu$ is such that only a single transverse mode, and hence two spin channels, are occupied. As this system is only quasi one-dimensional, rather than truly one-dimensional, it is difficult to extract analytical results. As a result, we look to integrate out the transverse dimension $y$ to construct an effective one-dimensional Hamiltonian $H^{\text{eff}}$ which can be analytically diagonalised. To this end, and following the work of Ref. \cite{parkAndreevSpinQubits2017}, the Hamiltonian is first partitioned into a free term $H_0$, and a perturbing term $V$
                \begin{equation}\label{eq::EffectiveHamiltonianPartition}
                    H_C = H_C^0 + V  \; ,
                \end{equation} 
            where the free term is given by
                \begin{equation}\label{eq::EffectiveHamiltonianPartitionFree}
                    H_C^0 =  -\frac{\hbar^{2}\nabla^{2}}{2m^*} - \mu  + i\alpha\sigma_y \partial_x +E_Z \sigma_y  \; ,
                \end{equation} 
            and the perturbation is given by
                \begin{equation}\label{eq::EffectiveHamiltonianPartitionPert}
                    V =  -i\alpha \sigma_x \partial_y \; .
                \end{equation} 
            As $H_C^0$ is both separable in $x$, $y$, and commutes with $\sigma_y$, then we can write the energy eigenstates as
                \begin{equation}\label{eq::EffectiveHamiltonianFreeEnergyEigenstates}
                    \Phi_{j\pm}(x,y) = e^{ik_xx}\phi_j(y)\chi_{\pm}  \; ,
                \end{equation} 
            where 
                \begin{equation}\label{eq::ConfinedStates}
                    \phi_{j\pm}(y) = \sqrt{ \frac{2}{W} }\sin\left( \frac{j\pi y}{W} \right)  \; ,
                \end{equation} 
            are orthogonal eigenstates of $-\hbar^{2}\partial_y^2 / (2m^*)$ that satisfy the hard-wall boundary conditions the confinement along the $y$-axis imposes. The eigenenergies of these transverse states are given by
                \begin{equation}\label{eq::TransverseEigenstates}
                    E_{j} = \frac{\hbar^{2}\pi^{2}j^{2}}{2mW^{2}}  \; .
                \end{equation} 

\begin{figure}[ht]
                    \centering
                    \includegraphics[width=0.5\textwidth]{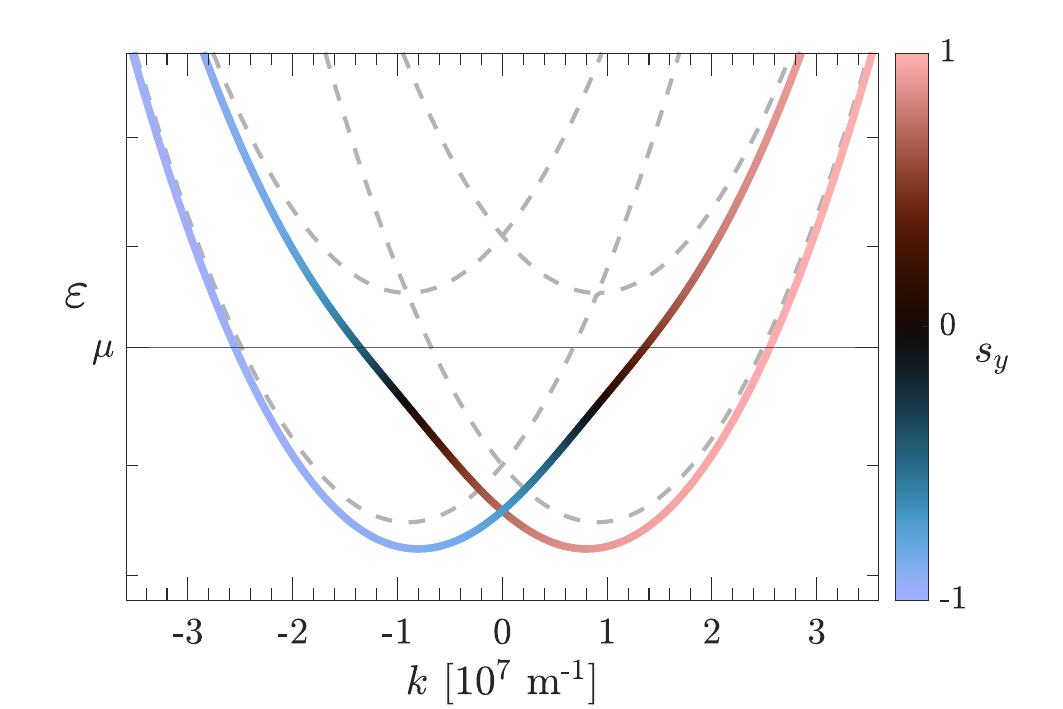}
                    \caption{Analytic dispersion relation computed using Eq. \ref{eq::EffectiveHamiltonianSpectrum} for a two-dimensional material with width $W = 300\,$nm, spin-orbit strength $\alpha = 3 \times 10^{-11}\SI{}{eV m}$, and no external magnetic field $E_Z = 0$. The dashed grey line is when $\eta = 0$ such that there is no coupling between adjacent transverse subbands -- we do not colour these lines as to simplify the figure. The qualitative similarity of this figure to the numerical bandstructure of Fig. \ref{fig::RashbaSpectrum}.b indicates the quality of the analytic dispersion relation.}
                    \label{fig::ExampleBans}
                \end{figure}

            To integrate out the $y$-dimension, a projection operator $P_0$ that projects onto the basis of $\phi_j$ is utilised. It is defined such that it takes a generic function $f(x,y)$ and expresses it in the basis of $\phi_j$:
                \begin{equation}\label{eq::EffectiveHamiltonianProjectionOperator}
                    P_0f(x,y) \equiv \sum_i \phi_i(y) \left[ \int \phi_i^*(y)f(x,y) \; \dd{y} \right] \; .
                \end{equation} 
            With the use of this projection operator, to first order in the perturbation the effective Hamiltonian is given by \cite{bravyiSchriefferWolffTransformation2011}
                \begin{equation}\label{eq::FirstOrderEffectiveHamiltonian}
                    H^{\text{eff}} \equiv P_{0}HP_{0}  \; .
                \end{equation} 
            Writing the resulting eigenvalue equation as
                \begin{equation}\label{eq::EigenvalueEuqation}
                    H^{\text{eff}}\Psi_{\pm}(x,y) = \varepsilon \Psi_{\pm}(x,y) \; ,
                \end{equation} 
            we can left multiply by $\phi_i^*$ and integrate over $y$ to obtain the matrix equation
                \begin{equation}\label{eq::SystemOfLinearEquations}
                    \left< H^{\text{eff}}\right>_{ij}\psi_{j\pm}(x) = \varepsilon \psi_{i\pm}(x)  \; ,
                \end{equation} 
            where 
                \begin{equation}\label{eq::ProjHam}
                    \left< H^{\text{eff}}\right>_{ij} \equiv \int \phi_i^* H^{\text{eff}} \phi_j \dd y  \; ,
                \end{equation} 
            and
                \begin{equation}\label{eq::ProjEig}
                    \psi_{i\pm}(x) \equiv  \int \phi_i^*\Psi_{\pm}(x,y) \dd y \; .
                \end{equation} 
            As only the lowest transverse subband is occupied, we can truncate the expansion of Eq. \ref{eq::SystemOfLinearEquations} to only include the first two transverse subbands. In doing so, the only term which mixes the transverse subbands in Eq. \ref{eq::ProjHam} is given by
                \begin{equation}\label{eq::ProjectionNonZero}
                        \begin{aligned}
                            \left< H^{\text{eff}}\right>_{12} = - \left< H^{\text{eff}}\right>_{21} &= -i\alpha\int \phi_1^* \sigma_x\partial_y \phi_2 \dd y  \\
                                                                          &= i\sigma_x\frac{8\alpha}{3W}\\
                                                                          &\equiv i\sigma_x\eta
                        \end{aligned} 
                \end{equation} 
            where $\eta \equiv 8\alpha / (3W)$ is the geometry dependent measure of the coupling between different transverse subbands of opposite spin. The diagonal terms are also given by
                \begin{equation}\label{eq::DiagonalTerms}
                    \begin{aligned}
                         \left< H^{\text{eff}}\right>_{11} &= \frac{\hbar^{2}k^2}{2m^*} + E_1 - \mu  -k\alpha\sigma_y +E_Z \sigma_y    \; ,\\
                         \left< H^{\text{eff}}\right>_{22} &= \frac{\hbar^{2}k^2}{2m^*} + E_2 - \mu  -k\alpha\sigma_y +E_Z \sigma_y    \; ,\\
                    \end{aligned} 
                \end{equation} 
            where again we are dropping subscripts to simply denote the momentum along the $x$-axis by $k$. Focusing on the lowest energy solution -- as only a single subband is assumed to be occupied -- the energy of the ground state $\varepsilon_{\pm}$ can be solved from Eq. \ref{eq::SystemOfLinearEquations}. In doing so, two bands are found with energy given by
                \begin{equation}\label{eq::EffectiveHamiltonianSpectrum}
                            \varepsilon_{\pm}(k) = \frac{\hbar^2k^2}{2m^*} + \frac{5E_1}{2}-\mu + \sqrt{\left(\frac{3E_1}{2} \pm E_Z \pm \alpha k \right)^2 + \eta^2} 
                \end{equation} 
            We plot an example dispersion relation using Eq. \ref{eq::EffectiveHamiltonianSpectrum} in Fig. \ref{fig::ExampleBans}. 
            Using Eq. \ref{eq::FermiVelocityDefinition} and Eq. \ref{eq::LinearisedSpinTextureDefinition}, the spin-projections and the group velocity as a function of the wavevector are given by \cite{parkAndreevSpinQubits2017}
            \begin{equation}\label{eq::SpinAndVel}
                    \begin{aligned}
                        s_{1\pm}(k) &=  \frac{\frac{3E_1}{2} \pm \alpha k}{ \sqrt{\left(\frac{3E_1}{2} \pm \alpha k\right)^2 + \eta^2}}   \; , \\
                        v_{1\pm}(k)  &= \frac{\hbar k}{m^*} + \frac{\alpha}{\hbar}\frac{\frac{3E_1}{2}\pm \alpha k}{ \sqrt{\left(\frac{3E_1}{2}\pm \alpha k\right)^2 + \eta^2}} \; .
                    \end{aligned}
            \end{equation} 
            The spin-projections $s_{1\pm}(k)$ are also plotted in Fig. \ref{fig::ExampleBans}. A useful diagnostic is that by setting $\eta = 0$, such that we are ignoring the overlap between adjacent transverse subbands, then Eq. \ref{eq::SpinAndVel} states that $s_{1 \pm} = \pm 1$ as expected.


            %
        \subsection{Anomalous phase}\label{ssec::SingleChannelAnomalouPhase}
            
            The general expression for the anomalous Josephson phase given by Eq. \ref{eq::AnomalousPhase} is written as a sum over all of occupied subbands; in systems with only a single occupied subband it can be simplified to
            \begin{equation}\label{eq::SubstitutedInAnomalousInSingleModeSystem}
                        \varphi_0  =   E_ZL \left(\frac{s^r_{1+}}{\hbar v^r_{1+}} + \frac{s^r_{1-}}{\hbar v^r_{1-}}\right)   \; .
            \end{equation} 
            If the asymmetry of the spin-texture is now ignored such that we naïvely simply set $s^{r}_{1+} = 1$ and $s^{r}_{1-} = -1$, then the anomalous phase is solely a function of the Fermi velocity asymmetry 
                \begin{equation}\label{eq::FakeSingleBandAnomalous}
                    \varphi_0   \underset{\substack{s^r_{1+}  = +1 \\ s^r_{1-} =-1}}{\longrightarrow}    E_ZL\left(\frac{1}{\hbar v^r_{1+}} - \frac{1}{\hbar v^r_{1-}}\right)   \; .
                \end{equation} 
            This equation is not new, having appeared in the literature previously to model the anomalous phase in quasi one-dimensional Josephson junctions \cite{kriveChiralSymmetryBreaking2004,yokoyamaAnomalousJosephsonEffect2014}. It is a good validation tool to note that ignoring the spin-texture our equation resolves to the standard result in the literature. 

\begin{figure}[ht]
                    \centering
                    \includegraphics[width=0.5\textwidth]{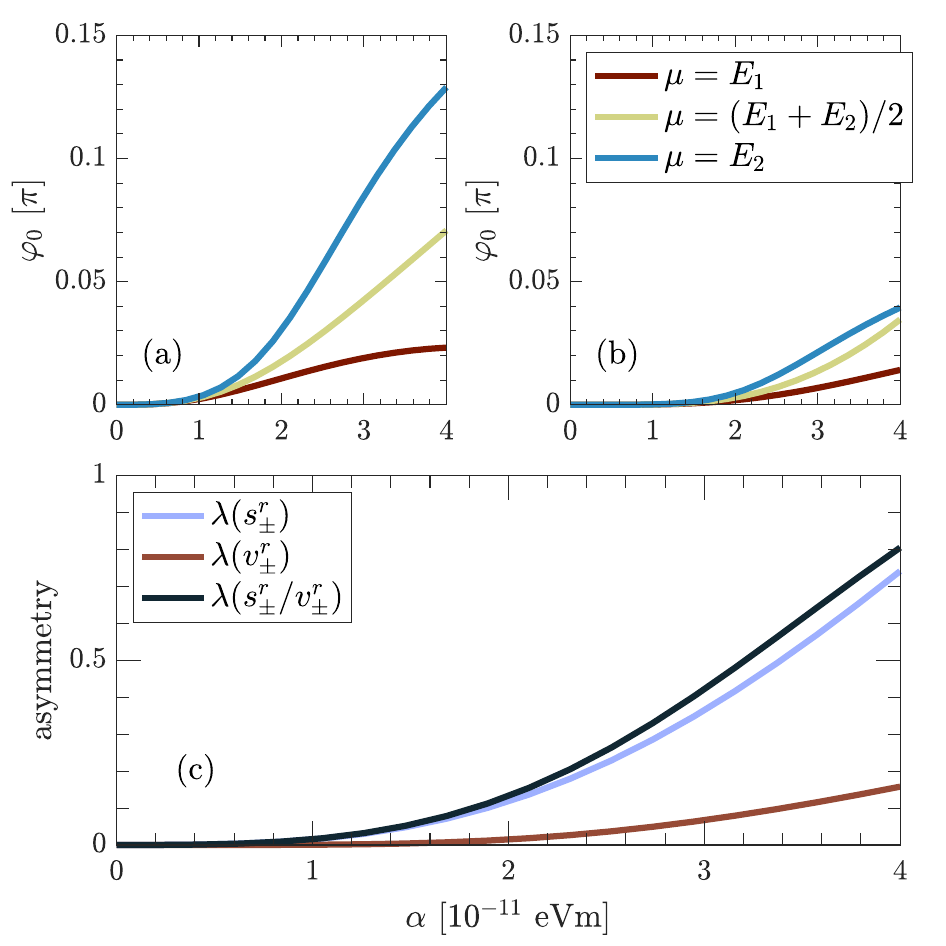}
                    \caption{The anomalous phase in the Josephson junction as a function of the spin-orbit coupling strength. As the effective one-dimensional model for the dispersion derived in Sec.\ref{ssec::EffectiveHamiltonian} is only correct to first order in the subband coupling, the spin-orbit strength is kept such that the energy within the spin-orbit energy scale does not exceed the subband energy scale $m^*\alpha^2 / \hbar^2 < E_1$. (a) The case of two-dimensional Rashba spin-orbit coupling with subband mixing. (b) The case ignoring the spin-texture, and considering only Fermi velocity asymmetry. (c) The asymmetry parameter computed using Eq. \ref{eq::asymParam}. The Fermi level $\mu$ has been fixed to the value $(E_1 + E_2)/2$ for each curve.}
                    \label{fig::totalShiftDiff}
                \end{figure}

            To highlight the explicit role of the spin-texture, we will now analytically solve for the anomalous phase using both Eq. \ref{eq::SubstitutedInAnomalousInSingleModeSystem}, which contains the spin-texture asymmetry, and the conventional Eq. \ref{eq::FakeSingleBandAnomalous}, which ignores the spin-texture. To this end, we consider a Josephson junction with a length of $200\,\text{nm}$, a width of $20\,\text{nm}$, and a magnetic field is applied along the $y$-axis with a Zeeman strength of $100\mu\text{eV}$. To compute the anomalous phase, the spin-projection and group velocities of the states at the Fermi level must be determined. To determine the Fermi wavevectors $k_{1\pm}$, which can then be substituted into Eq. \ref{eq::SpinAndVel} to determine the spin-projection and Fermi velocities, the roots of the effective dispersion relation derived in Eq. \ref{eq::EffectiveHamiltonianSpectrum} must be found. Although this is equivalent to solving for the roots of a quartic, and hence has an analytic form, it is not illuminating to write down the resulting expressions as they are particularly cumbersome. Regardless, substituting in these roots into Eq. \ref{eq::SpinAndVel} provides an analytic form for the Fermi velocity and spin-projection at the Fermi level. Substituting these into Eq. \ref{eq::SubstitutedInAnomalousInSingleModeSystem} and Eq. \ref{eq::FakeSingleBandAnomalous} will then provide analytic forms for the anomalous phase. Sweeping both the strength of the Rashba spin-orbit strength $\alpha$ and the Fermi level $\mu$, the results are plotted in Fig. \ref{fig::totalShiftDiff}.a and Fig. \ref{fig::totalShiftDiff}.b respectively.

            By comparing  Fig. \ref{fig::totalShiftDiff}.a and  Fig. \ref{fig::totalShiftDiff}.b, it is clear that the case with the spin-texture results in a significantly larger anomalous phase. The equation for anomalous phase in Eq. \ref{eq::SubstitutedInAnomalousInSingleModeSystem} indicates that the driver of the anomalous phase is the asymmetry between the two spin-bands of the system. To this end, we define an asymmetry parameter for some quantity $x_{\pm}$ as \cite{kriveChiralSymmetryBreaking2004}
                \begin{equation}\label{eq::asymParam}
                    \lambda(x_{\pm})\equiv \abs{\frac{x_+ - x_-}{x_+ + x_-}}  \; .
                \end{equation} 
            Figure \ref{fig::totalShiftDiff}.c compares the asymmetry of $s^r_{1\pm}$ with $v^r_{1\pm}$ and their ratio $s^r_{1\pm} / v^r_{1\pm}$. Fig. \ref{fig::totalShiftDiff}.c indicates that the asymmetry within the device is almost entirely due to the spin-texture rather than the Fermi velocity asymmetry. This is not unsurprising as although increasing the spin-orbit coupling causes the Fermi velocities to deviate slightly, the spin-projections can experience a full sign change. These results indicate that it is the spin-texture which is the largest driver of the anomalous Josephson effect.

            
    %
%

%
\section{Anomalous phase in multi-channel systems}\label{sec::MultiAnomalous}

    In Sec. \ref{sec::AnomalousPhaseSingle}, having only a single subband occupied allowed for a number of analytic simplifications. However, in systems containing multiple occupied subbands, and with a two-dimensional Rashba spin-orbit term mixing states, it is not possible to evaluate the dispersion relation analytically. As the generic expression for the anomalous phase in Eq. \ref{eq::AnomalousPhase} requires the spin-projection and group velocity of every state at the Fermi level, then it is also not possible to derive analytic expressions for the resulting anomalous phase. Fortunately, there are some approximations which can be made when multiple transverse subbands are occupied. For example, for a large enough Fermi level relative to the subband spacing, the linearised dispersions will appear almost parallel. As a result, we can replace the Fermi velocity of each state by simply the velocity of the lowest subband $v_1$; this is a reasonable upper bound as in the absence of spin-orbit interactions this would be the supremum of the set of Fermi velocities.
    %
    By making this replacement, the phase gained by the Andreev bound state moving through the magnetic field, defined in Eq. \ref{eq::PhaseQuantisationCOndition}, is now given by
        \begin{equation}\label{eq::PhaseQuantisationCOnditionNew}
            \frac{2Ls_{j\pm}^{r}E_{Z}}{\hbar v_{j\pm}^{r}}  \to \frac{2Ls_{j\pm}^{r}E_{Z}}{\hbar v_1} \equiv \theta_Bs_{j,\pm}^{r}   \; ,
        \end{equation} 
    where we have pulled out all constant coefficients into a magnetic field induced phase factor $\theta_B$ \cite{yokoyamaAnomalousJosephsonEffect2014}. This simplification explicitly ignores Fermi velocity asymmetry, allowing us to focus entirely on the spin-texture asymmetry; even for the case of a single occupied band, we demonstrated in Fig. \ref{fig::totalShiftDiff}.c that this is a reasonable simplification to make. 
    This can then be substituted into the expression for the total phase shift given by Eq. \ref{eq::AnomalousPhase} 
    \begin{equation}\label{eq::SimpleExpression}
        \varphi_0   = \atan \left[\frac{\sum_{j} \left( \sin \theta_B s_{j+}^{r}   +  \sin \theta_B s_{j-}^{r}  \right)}{\sum_{j} \left( \cos \theta_B s_{j+}^{r}   +  \cos \theta_B s_{j-}^{r}  \right)}\right]  \; .
    \end{equation} 

    Another approximation we can make is that in multi-channel systems, the Fermi wavelength of the electron within the lowest subband is significantly smaller than the confined width of the system $\lambda_1 \ll W$. Combining this with the fact that we are considering systems where the Zeeman energy is smaller than the subband spacing, $E_Z < \Delta E \sim \hbar^2 / (2mW^2)$, then we can bound the magnetic phase $\theta_B$ by
        \begin{equation}\label{eq::ThetaBExp}
                \begin{aligned}
                    \theta_B \equiv  \frac{2LE_{Z}}{\hbar v_{1}^{r}} < \frac{2L\Delta E}{\hbar v_{1}^{r}} \sim \frac{L}{W}\frac{\lambda_{1}}{W} \; ,
                \end{aligned} 
        \end{equation} 
    where we have substituted in the Fermi wavelength of the lowest subband electron $\lambda_{j} \equiv h / mv_j$. By making the multi-channel approximation that $\lambda_1 \ll W$, and implicitly assuming that the length of the device is not significantly greater than its width, we find that $\theta_B \ll 1$. Utilising this result, the anomalous phase in Eq. \ref{eq::SimpleExpression} can be Taylor expanded to first order in the magnetic field strength to write
    \begin{equation}\label{eq::FirstOrderExpansionAnomalousPhaseTwo}
            \begin{aligned}
                \varphi_0   &\sim \atan \left[\frac{\theta_B}{2N}\sum_j^N \left(  s_{j+}^{r} + s_{j-}^{r} \right) \right]      \; , \\
                &\sim  \frac{\theta_B}{2N}\sum_j^N \left(  s_{j+}^{r} + s_{j-}^{r} \right)      \; , \\
                &\equiv \left< s_y \right>\theta_B \; ,
            \end{aligned} 
    \end{equation} 
    where we have defined the average spin-projection over all $N$ occupied subbands as
        \begin{equation}\label{eq::DefinitionAverage}
            \left< s_y \right> \equiv  \frac{1}{2N}\sum_j^N \left(  s_{j+}^{r} + s_{j-}^{r} \right)    \; .
        \end{equation} 
    Eq. \ref{eq::FirstOrderExpansionAnomalousPhaseTwo} is one of the main results of this work -- namely, an expression for the anomalous phase which is linear in the external magnetic field, and directly proportional to the spin-texture asymmetry. Importantly, the only physical mechanism used to derive this result was the inclusion of the spin-texture induced by the two-dimensional Rashba spin-orbit coupling. 

    \subsection{Average spin-projection in multi-channel systems}\label{ssec::BandStructure2D}
        
\begin{figure}[ht]
                \centering
                \includegraphics[width=0.5\textwidth]{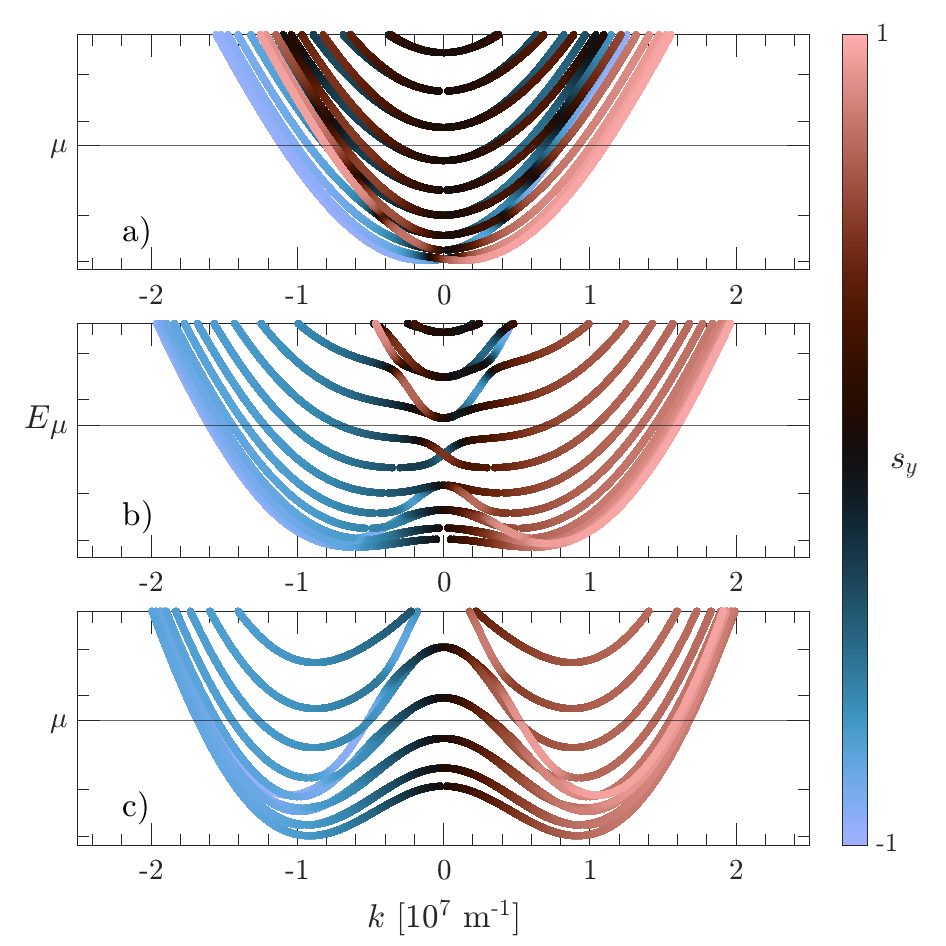}
                \caption{Dispersion relation for a system with seven occupied subbands -- totalling fourteen different spin states -- with a two-dimensional Rashba spin-orbit coupling term and no external magnetic field. In each plot the spin-orbit coupling strength is set such that: (a) $k_\alpha = 0.1k_1$ (b) $k_\alpha = k_1$ (c) $k_\alpha = 2k_1$. Note that unlike Fig. \ref{fig::RashbaSpectrum}, here we are considering a wide range of spin-orbit strengths}
                \label{fig::RashbaShift}
            \end{figure}

        To understand the microscopic origin that gives rise to a non-zero average spin-projection, $\left<s_y\right> \neq 0$, it is worth considering some generic features of the transverse subbands of the non-superconducting region of the Josephson junction. Indeed, we can view the two-dimensional Rashba spin-orbit coupling as one term, $\alpha\sigma_yp_x$,  which shifts the wavevectors of the different spin-states by some amount $\pm k_\alpha$, where \cite{manchonNewPerspectivesRashba2015}
            \begin{equation}\label{eq::FermiWavevectorDef}
                k_{\alpha} = \frac{m\alpha}{\hbar^2}  \; ,
            \end{equation} 
        and another term, $\alpha\sigma_xp_y$, which couples the transverse subbands and mixes the spins. To illustrate this, the dispersion relation for a system with $N=7$, no magnetic field, and varying spin-orbit strengths is plotted in Fig. \ref{fig::RashbaShift}. As in Fig. \ref{fig::RashbaSpectrum}, this was computed using the tight-binding calculation described in Ref. \cite{mirelesBallisticSpinpolarizedTransport2001}, however in this case we are considering a wider range of spin-orbit values. The output eigenstates of this numerical calculation can then be used to extract the spin-projection of that state by solving Eq. \ref{eq::LinearisedSpinTextureDefinition} numerically.

        From analysing Fig. \ref{fig::RashbaShift}, we observe that even for small spin-orbit strengths there is significant mixing between the spin-projections of right-moving states. However, when $k_\alpha\sim k_1$ the coupling between the different spin states is great enough that the spin-projection of each state is locked to the sign of its wavevector i.e. in this limit essentially every right-moving state is spin-up. This is strongly reminiscent of the band-structure of one-dimensional nanowires in the topological insulator limit where instead of a two dimensional spin-orbit interaction, a magnetic field parallel to the nanowire provides the coupling between spin states \cite{oppenTopologicalSuperconductingPhases2017}. Interestingly, when the spin-orbit strength is increased such that $k_\alpha \gg k_1$, the different spin bands are so far shifted from each other that they effectively decouple. This results in two almost separate parabolas which appear to have no subband coupling. As a result, the right-moving states are either well-defined spin-up or spin-down states. 

\begin{figure}[ht]
                \centering
                \includegraphics[width=0.5\textwidth]{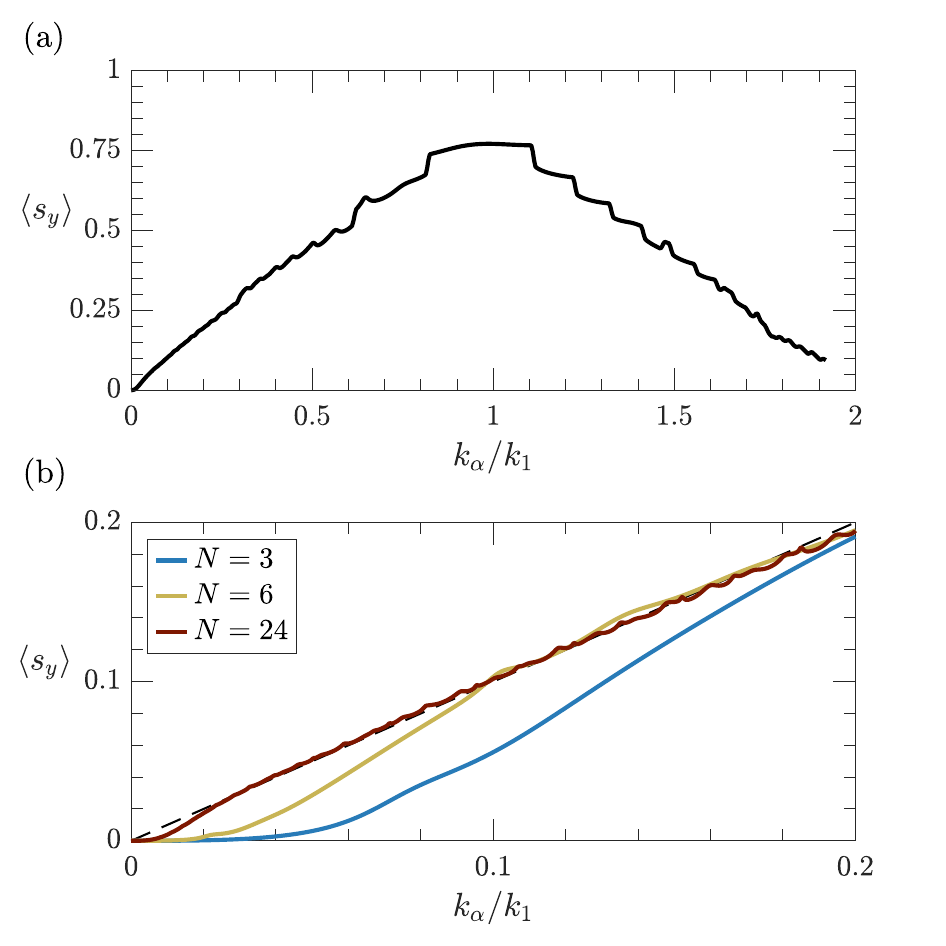}
                \caption{(a) Average spin-projection $\left<s_y\right>$ for the system defined in Fig. \ref{fig::RashbaShift} as a function of the Rashba spin-orbit coupling strength. (b) The average spin-projection of three systems where the number of occupied subbands differs -- every other parameter is kept fixed. The spin-orbit wavevector has been restricted to $0 \leq k_\alpha \leq 0.2k_1$ such that we are best observing only the `linear' region. }
                \label{fig::SpinExpect}
            \end{figure}

        To investigate this point further, by considering the same system as in Fig. \ref{fig::RashbaShift}, we can extract the spin-projections at the Fermi level to compute the average spin-projection. We plot the corresponding average spin-projection in Fig. \ref{fig::SpinExpect}.a. The net result is a function that initially increases somewhat linearly before reaching a maximum at the point $k_\alpha = k_1$, then decreasing to effectively zero at $k_\alpha = 2k_1$. The sharp jumps observed in the average are due to discrete changes in the number $N$ of right-moving states within the system which strongly perturbs the mean. 

        These qualitative features can be understood from the dispersion relation for the states depicted in Fig. \ref{fig::RashbaShift}. When $k_\alpha \sim k_1$ we expect to observe a maximum average spin-projection as all right-moving states have positive spin-index. Similarly, when the spin-orbit strength is increased further, the spin bands effectively decouple such that the spin-projections of the occupied spin-up and spin-down states cancel -- hence the average should decreases towards zero as observed.

        A particular feature of Fig. \ref{fig::SpinExpect}.a which we want to focus on is that for $0 \leq k_\alpha \ll k_1$, the average spin-projection appears to be a linear function of the spin-orbit coupling strength
            \begin{equation}\label{eq::AverageSpin}
                \left< s_y \right> \sim \frac{k_\alpha}{k_1} \; .
            \end{equation} 
        To understand this, we first note that what perturbs the average spin-projection is not the mixing between occupied states -- as mixing between occupied states does not change the sum of the occupied spin-projections due to pairwise cancellation -- but rather unoccupied higher subbands which mix with an occupied state. This is most clear in the case of a single occupied subband studied in Sec. \ref{ssec::EffectiveHamiltonian}, where it was the coupling with the unoccupied $j=2$ subband with the occupied $j=1$ subband which drove the effect. We leave the full description for Appendix. \ref{sec::LinearAverageSpin}, however, the reason for the apparent linear increase is that the highest occupied spin-down state has its spin fully flipped whenever the spin-orbit strength wavevector $k_\alpha$ is an integer multiple of $k_1 / N$. Effectively, as a function of the spin-orbit strength, the total sum of the spin-projections increases by two at regular intervals of $k_1 / N$. In the limit as $N\to\infty$, this leads to the apparent linear increase in the average spin-projection. To demonstrate this, in Fig. \ref{fig::SpinExpect}.b we plot the average spin-projection for systems with different numbers of occupied states $N$ and observe that the average spin-projection progressively becomes more linear with $N \to \infty$.

    \subsection{Anomalous phase}\label{ssec::AnomalousPhaseMulti}
        The expression for the anomalous phase in multi-channel systems given by Eq. \ref{eq::FirstOrderExpansionAnomalousPhaseTwo} states that the anomalous phase is equal to the average spin-projection multiplied by the phase factor induced by the magnetic field $\theta_B$. From the results of Sec. \ref{ssec::BandStructure2D}, this allows us to make some qualitative predictions on the resulting anomalous phase. Most interestingly, for Rashba spin-orbit strengths such that $k_\alpha < k_1$, the resulting average spin-projection can be written approximately as $ \left< s_y \right> \approx k_\alpha/k_1$. As a result, within this regime the total anomalous phase is given by the formula 
        \begin{equation}\label{eq::ResultingAnomalousPhaseLinear}
            \varphi_0 = \frac{k_\alpha\theta_B}{k_1}  \; .
        \end{equation} 
        By substituting in the relation between the Fermi wavevector and velocity $k_1 \sim mv_1 / \hbar$, and the expression for $\theta_B$ given in Eq. \ref{eq::PhaseQuantisationCOnditionNew}, then the anomalous phase is given by
        \begin{equation}\label{eq::ResultingAnomalousPhaseBuzdin}
            \varphi_0 = \frac{2\alpha E_Z L}{\hbar^2 v_1^2}  \; .
        \end{equation} 
        Up to a factor of two, Eq. \ref{eq::ResultingAnomalousPhaseBuzdin} is the same expression as derived in Ref. \cite{buzdinDirectCouplingMagnetism2008} within the context of ballistic junctions in the long junction limit $L \gg \xi$. In that work, the model was not microscopic but instead phenomenological hence it is interesting to note that the results are in such good agreement. 

        More generically, as the spin-orbit strength is increased further such that $\alpha \sim \hbar v_1$ then the average spin-projection reaches a maximum value close to unity such that 
        \begin{equation}\label{eq::ResultingAnomalousPhaseSat}
            \varphi_0 \sim \theta_B \; ,
        \end{equation} 
        where small changes in the spin-orbit strength have little to no effect on the resulting anomalous phase. Increasing the spin-orbit strength further such that $\alpha \gg \hbar v_1$, the average spin-projection should appear to decrease linearly as the different spin bands decouple. These heuristic expectations on the anomalous phase can be conveyed by the piecewise function
            \begin{equation}\label{eq::PiecewiseAnom}   
                \varphi_0 \sim \begin{cases}
                    \theta_B  {\alpha}/({\hbar v_1})\; ,  & \alpha \ll \hbar v_1 \\
                    \theta_B \; , & \alpha \sim \hbar v_1 \\
                    \theta_B \left[2 - {\alpha}/({\hbar v_1})\right]\; ,   & \alpha \gg \hbar v_1 \\
                    0 \; ,&\alpha \to \infty
                \end{cases}  \; 
            \end{equation} 
\section{Numerical results}\label{sec::NumericalResults}

    Having developed a mechanism for the anomalous Josephson effect specifying the role of the spin texture, we now look to compare the predictions with numerical calculations which solve for the anomalous phase directly. To do so the Hamiltonian of Eq. \ref{eq::CentralHam} is cast into a tight binding model and the anomalous phase is solved for using the non-equilibrium Green's function formalism (NEGF). 
    
    For the remainder of this section the electron effective mass $m^*$ is set to $0.023m$, and the temperature set such that $2k_B T = |\Delta|$ and hence, as usual, we are working close to the critical temperature; both are assumed constant throughout the system. The magnitude of the superconducting order parameter is set to $1\mu$eV -- although this is reasonably small, it is simply to enforce the short junction limit for every numerical result and does not qualitatively change the results. Similarly, a temperature close to the critical temperature was used simply to retain a sinusoidal Josephson current which best illustrates the anomalous phase shift.

    \subsection{Numerical methods}\label{ssec::NumericalMethods}
        We consider a discrete system with lattice spacing $a$ and a length of $Na$ and a width of $Ma$ for integers $N$ and $M$. Within a tight-binding model, the Hamiltonian for the non-superconducting central region of the device, given by Eq. \ref{eq::CentralHam}, is written as
            \begin{equation}\label{eq::TightBindingNonSuper}
                    \begin{aligned}
                        H_C =  &\sum_{ (i,j) \in C } \psi_{i,j}^{\dagger} \left[ 4t + E_Z \sigma_y - \mu \right]\psi_{i,j}  \\
                       &-\sum_{\langle i,j\rangle }\left[\psi_{i,j}^{\dagger} \left( t +\frac{i\alpha}{2a}\sigma_y \right)\psi_{i+1,j}\right.\\
                        &\qquad + \left. \psi_{i,j}^{\dagger} \left( t -\frac{i\alpha}{2a}\sigma_x \right)\psi_{i,j+1}+ \text{h.c.} \right] \; ,
                    \end{aligned} 
            \end{equation} 
        where the operator $\psi_{i,j} \equiv \begin{pmatrix}
            \psi_{i,j;\uparrow} & \psi_{i,j;\downarrow}
        \end{pmatrix}^T$ annihilates an electron at the lattice site $(i,j)$ \cite{jauhoTimedependentTransportInteracting1994,martin-roderoMicroscopicTheoryJosephson1994,liuAnomalousJosephsonCurrent2010}. The hopping parameter $t\equiv \hbar / (2m^*a^2)$ factors in both the effective mass $m^*$ and the lattice spacing $a$ of the simulation. Written in this quadratic form, we can extract the corresponding first quantised Hamiltonians for the central region as an $(2NM \times 2NM)$ block tri-diagonal matrix 
            \begin{equation}\label{eq::TriDiagFull}
                h_C = \begin{bmatrix}
                    \varepsilon_1   & t_1 &&&\\
                    t_1^{\dagger} &\varepsilon_2   & t_2 &&\\
                    &\ddots &\ddots&\ddots&&\\
                    &&t_{N-1}^{\dagger} &\varepsilon_{N-1}   & t_N\\
                    &&& t_N^\dagger & \varepsilon_N
                \end{bmatrix}  \; ,
            \end{equation} 
        where
        \begin{equation}\label{eq::TriDiagSubSpace}
                \begin{aligned}
                    \varepsilon_j &= \text{TriDiag}\left[-t+\frac{i\alpha}{2a}\sigma_y , \, 4t  - \mu + E_Z\sigma_y, \, -t-\frac{i\alpha}{2a}\sigma_y  \right]  \; , \\
                    t_j &= \text{Diag}\left[-t + \frac{i\alpha}{2a}\sigma_x \right]  \; ,
                \end{aligned} 
        \end{equation} 
        are themselves $(2M \times 2M)$ banded matrices \cite{sriramSupercurrentInterferenceSemiconductor2019}.

        For the superconducting sections, as indicated in Eq. \ref{eq::LeftRightFull}, the degrees of freedom are doubled to include the holes within the system
            \begin{equation}\label{eq::NambuSpinor}
                \Psi =  \begin{pmatrix}
                    \psi \\ \psi^{*}
                \end{pmatrix}  \; ,
            \end{equation} 
        where $\psi$ is a vector containing the annihilation operator at every lattice site. The superconducting order parameter $\Delta$ now acts to couple these two subspaces \cite{zengElectronicTransportHybrid2003} 
            \begin{equation}\label{eq::TightBindingNonSuper}
                    \begin{aligned}
                        H_{L/R} &=  \Psi^{\dagger}h_{L/R} \Psi\\
                      &=  \begin{pmatrix}
                            \psi^{\dagger} & \psi^{T}
                        \end{pmatrix} \begin{pmatrix}
                        h_0 + h_{\alpha} & i\sigma_y\Delta_{L/R}\\
                            -i\sigma_y\Delta_{L/R}   & -( h_0 + h_{\alpha})^* 
                        \end{pmatrix}  \begin{pmatrix}
                            \psi \\ \psi^{*}
                        \end{pmatrix}  
                    \end{aligned} 
            \end{equation} 

        Using the discretised first-quantised Hamiltonians the Green's functions for the central region at energy $\varepsilon$ can be determined by solving the corresponding Dyson's equation \cite{rammerQuantumFieldTheory2007}
            \begin{equation}\label{eq::DysonEquation}
            \mathcal{G}(\varepsilon) = [(\varepsilon + i\kappa) \mathbb{I} - h_C - \Sigma_L - \Sigma_R]^{-1}  \; ,
            \end{equation} 
        where $\Sigma_{L/R}$ are the self-energy terms due to the coupling from the leads, and $\kappa$ is an infinitesimal energy perturbation. By assuming that the leads are spatially homogenous along the $x$-axis, then the self-energies can be solved using an efficient recursive process \cite{sanchoQuickIterativeScheme1984}. Furthermore, assuming both leads are at thermal equilibrium at temperature $T$, a necessary quantity known as the lesser self-energy, $\Sigma^<$, can be computed as \cite{jauhoTimedependentTransportInteracting1994}
            \begin{equation}\label{eq::LevelBroadening}
                \begin{aligned}
                    \Sigma^< &=f(\varepsilon)(\Sigma_L^{\dagger}-\Sigma_L) + (\Sigma_R^{\dagger}-\Sigma_R)f(\varepsilon) \; ,\\
                        &\equiv   \Sigma^<_L + \Sigma^<_R 
                \end{aligned} 
            \end{equation} 
        where $f(\varepsilon)$ is the usual Fermi function.
        
        From these quantities, the current can be computed by \cite{jauhoTimedependentTransportInteracting1994}
            \begin{equation}\label{eq::CurrentFromLead}
                I = e\int \left(\mathcal{G}\Sigma^<_L +   \mathcal{G}\Sigma^<\mathcal{G}^{\dagger}\Sigma_L^\dagger\right) \dd \omega\; .
            \end{equation} 
        Another important quantity is the density of states within the device which is computed by \cite{rammerQuantumFieldTheory2007}
            \begin{equation}\label{eq::DOS}
                D = i\Tr\left[\mathcal{G} - \mathcal{G}^\dagger\right]  \; ,
            \end{equation} 
        where the trace is over both spatial and spin indices. 
    \subsection{Single-mode system}\label{ssec::SingleModeNumerics}

        In Sec. \ref{sec::AnomalousPhaseSingle} analytic formulas for the anomalous phase shift in single band systems were derived. We now want to explicitly evaluate the anomalous phase numerically for these single-band systems to compare with the analytic formulas. To do so, the Fermi level $\mu$ of the numerical simulations will be tuned such that it lies in-between the first and second transverse subband energy. To keep the electron density constant as the Rashba spin-orbit interaction is increased, the Fermi level is modified as \cite{yokoyamaAnomalousJosephsonEffect2014}
            \begin{equation}\label{eq::ModificationFermiLevel}
                \mu \to \mu - \frac{\hbar^2 k_\alpha^2}{2m^*}  \; .
            \end{equation} 
        The lattice spacing was set to $0.1\,\mu$m, and the width and length to $0.8\,\mu$m and $12\,\mu$m respectively. It is important to reiterate that this is not a one-dimensional simulation, but rather a two-dimensional simulation with a tuned Fermi level such that only a single band is occupied.

\begin{figure}[ht]
                \centering
                \includegraphics[width=0.5\textwidth]{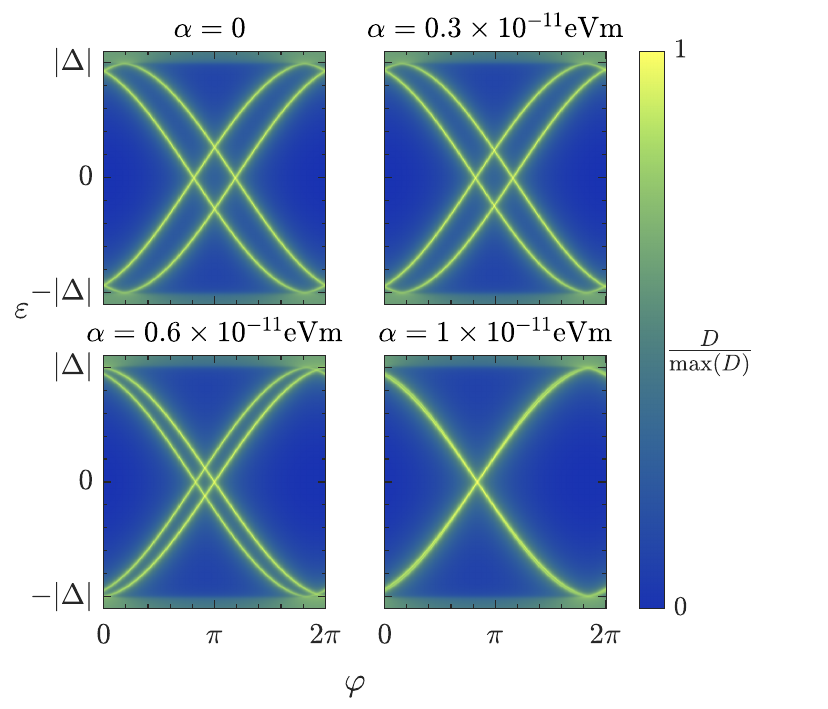}
                \caption{Normalised density of states computed using Eq. \ref{eq::DOS} for a Josephson junction with width $800\;$nm and length $12\mu$m as a function of both energy and phase difference $\varphi$ across the superconducting leads. The Fermi level $\mu$ has been tuned such that only a single subband is occupied. As stated in the text, a small magnetic field with Zeeman strength of $E_Z = 15\,\mu$eV is applied to induce a phase shift between the Andreev bound states. As the Rashba spin-orbit coupling strength $\alpha$ is increased the asymmetry between the two spin-bands becomes evident.}
                \label{fig::singleBandABS}
            \end{figure}

        To confirm that only a single subband is occupied, the normalised density of states $D$ as a function of the phase difference $\varphi$ across the Josephson junction is plotted in Fig. \ref{fig::singleBandABS}. In each plot the spin-orbit coupling strength $\alpha$ is varied; to break the spin degeneracy, a constant magnetic field with strength $E_Z = 15\,\mu$eV is also applied. 
        We can compare the peaks of the density of states with the equations for the Andreev bound state energies given by Eq. \ref{eq::FinalAndreevEnergy}. For the case of no spin-orbit coupling ($\alpha=0$) such that the states have trivial spin-projection, then the magnetic field identically shifts each Andreev bound state in the opposite direction by an amount given by $\theta_B$ -- this is essentially the process depicted in Fig. \ref{fig::ABSdiagram}.b. However, increasing the strength of the Rashba spin-orbit interaction leads to a mixing of the spin-projections. This causes the amount each band is shifted to become asymmetric -- a process illustrated in Fig. \ref{fig::ABSdiagram}.c. This is taken to the extreme when $\alpha = 1\times 10^{-11} $eVm where both Andreev bound states are visually coincident. This indicates that at this spin-orbit coupling strength, the formerly spin-down state is so strongly coupled with the higher order spin-up subband that its spin has effectively flipped. Due to this spin-projection flipping, both occupied states behave identically under the external magnetic field. Note that this change in direction of the phase shift cannot be explained solely by Fermi velocity asymmetry, as the sign of the Fermi velocity of the right-moving state is always positive.

            \begin{figure}[ht]
                \centering
                \includegraphics[width=0.5\textwidth]{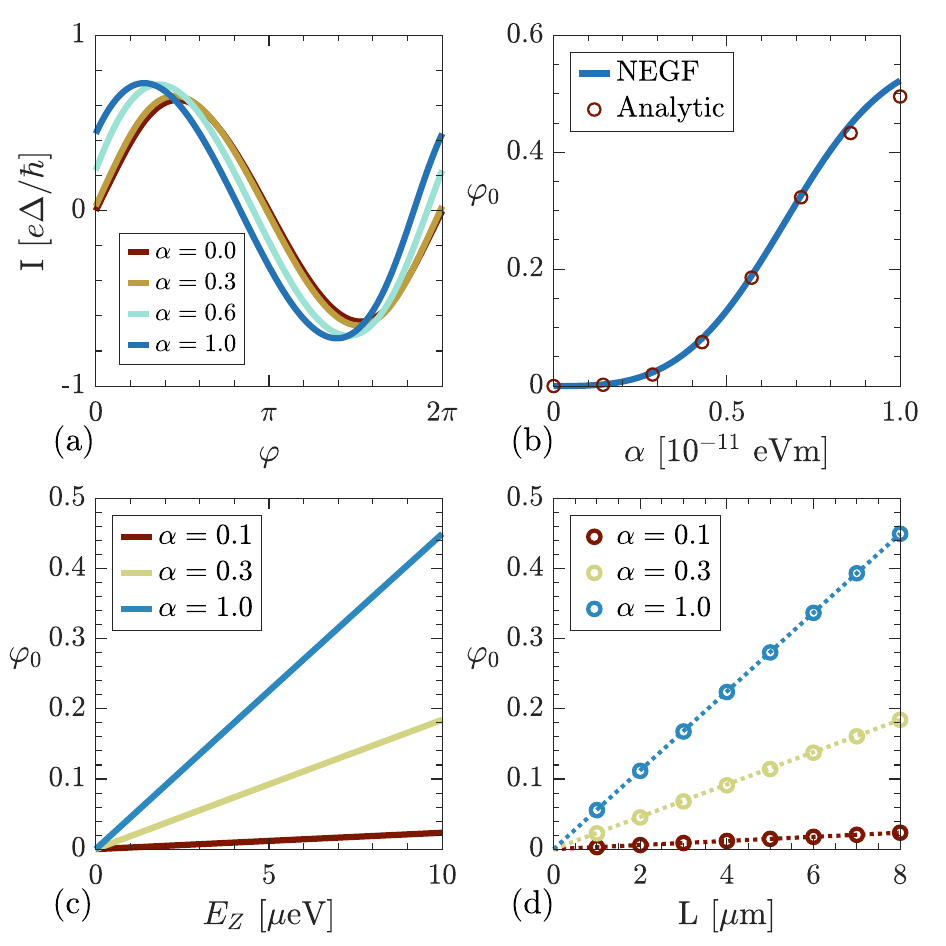}
                \caption{(a) The Josephson current corresponding to Fig. \ref{fig::singleBandABS} evaluated using Eq. \ref{eq::CurrentFromLead}. (b) The extracted anomalous phase for the numerical NEGF calculation overlaid over the results from the analytic band structure calculation given by solving Eq. \ref{eq::SubstitutedInAnomalousInSingleModeSystem}. (c) The anomalous phase as a function of the external magnetic field. (d) The anomalous phase as a function of the length $L$ of the system. As the length is quantised in units of the lattice spacing $a$, we only plot a discrete set of data points; unlike in (c) where a much higher resolution could be simulated. Note that the ratio between the magnetic field $B$ and the length $L$ of the device was held fixed which explains why the plots (c) and (d) appear identical. In each plot the Rashba spin-orbit coupling strength has units of $10^{-11}$eVm -- this information was removed to safe space within the figure. }
                \label{fig::NEGFSingleBandCurrent}
            \end{figure} 

        The Josephson currents corresponding to each trace of Fig. \ref{fig::singleBandABS} are plotted in Fig. \ref{fig::NEGFSingleBandCurrent}.a by solving Eq. \ref{eq::CurrentFromLead} numerically. Fig. \ref{fig::NEGFSingleBandCurrent}.a clearly demonstrates that as the spin-orbit coupling strength is increased, the sinusoidal Josephson current experiences a phase shift. Although not the purpose of this work, we also observe that the critical current -- the maximum supercurrent -- increases slightly with the spin-orbit coupling strength, as predicted in Ref. \cite{kriveChiralSymmetryBreaking2004}. This anomalous phase shift can then be extracted and plotted as a function of the Rashba spin-orbit strength, which is shown in Fig. \ref{fig::NEGFSingleBandCurrent}.b. Overlaid on top is the analytic result for the anomalous phase shift given by Eq. \ref{eq::SubstitutedInAnomalousInSingleModeSystem}. For small spin-orbit strengths the agreement between the two results are excellent, however, for larger spin-orbit strengths a slight discrepancy is visible; this is likely due to the fact that the analytic result was only computed perturbatively up to first order in the spin-orbit strength. 

        Another computational test for the analytic equation given by Eq. \ref{eq::SubstitutedInAnomalousInSingleModeSystem} is by numerically evaluating the anomalous phase as a function of both the Zeeman strength $E_Z$, and the length $L$ of the device. The results for this are plotted in Fig. \ref{fig::NEGFSingleBandCurrent}.c and Fig \ref{fig::NEGFSingleBandCurrent}.d respectively. The plots here are strikingly linear, which is unsurprising given that the numerical simulations are strictly within the short-junction limit and that the Zeeman energy is significantly smaller than the subband spacing -- the two requirements for the linear behaviour discussed in Sec. \ref{ssec::ABS}. 
        
    %

    %
    \subsection{Multi-mode system}\label{ssec::MultiModeNumerics}
        
        In this section, we want to compare with the results of Sec. \ref{sec::MultiAnomalous} concerning systems with multiple occupied subbands. To do so numerically, the lattice spacing $a$ is set to $0.25\mu$m and the width and length to $10\, \mu$m and length $20\, \mu$m respectively. Although this constitutes a rather large device, the small subband spacing this induces makes the numerical calculations easier. It should be noted that although this small subband spacing imposes a strict energy-scale for the other parameters such that we can observe the anomalous physics -- on the order of $\mu$eV -- it does not qualitatively change the results. 
        
\begin{figure}[ht]
                \centering
                \includegraphics[width=0.5\textwidth]{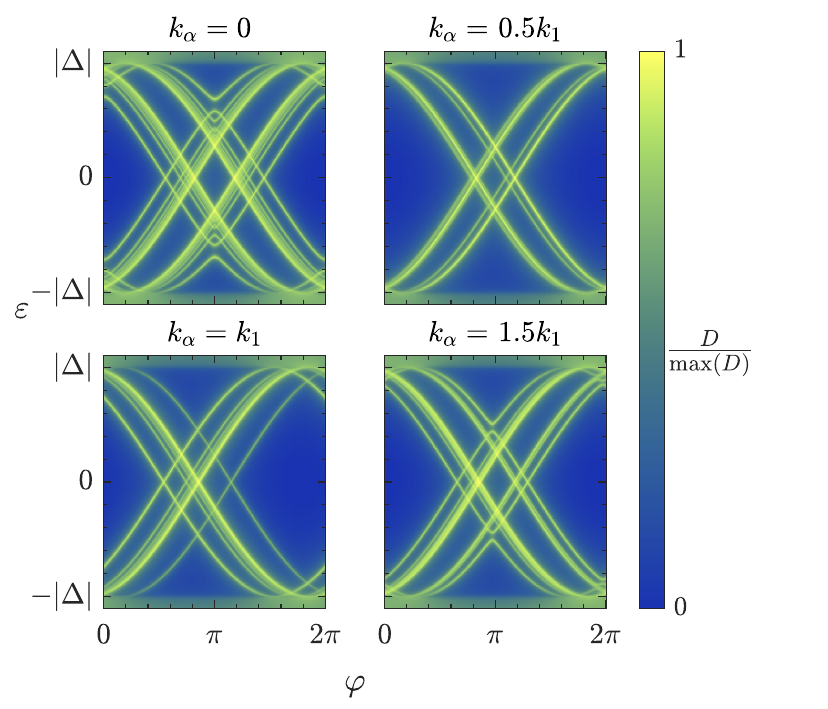}
                \caption{Normalised density of states computed using Eq. \ref{eq::DOS} for a Josephson junction with width $10\mu$m and length $20\mu$m as a function of both energy and phase difference $\varphi$ across the superconducting leads. The Fermi level is tuned such that there are twelve occupied subbands. As stated in the text, a small magnetic field of Zeeman strength $E_Z = 15\;\mu$eV is applied to induce a relative phase shift between the Andreev bound states. }
                \label{fig::multiBandABS}
            \end{figure}

        We tune the Fermi level such that there are twelve occupied states ($N=12$). As in the single-band case, to illustrate that there are multiple occupied states, we can plot the density of states for the Josephson junction as a function of the phase difference $\varphi$ across the two superconductors. As before, to break the spin degeneracy of the system a magnetic field with strength $E_Z = 15\mu$eV is applied. This result is illustrated in Fig. \ref{fig::multiBandABS}. Note that rather than referring to the spin-orbit strength $\alpha$ directly, it is simpler to refer to the spin-orbit wavevector $k_\alpha$ defined in Eq. \ref{fig::NEGFSingleBandCurrent}.

        Making the multi-channel approximations discussed in Sec. \ref{ssec::BandStructure2D}, the energy of each Andreev bound states derived in Eq. \ref{eq::FinalAndreevEnergy} can be written as
        \begin{equation}\label{eq::FinalAndreevEnergyTheta} 
             \begin{aligned}
                 \varepsilon^{r}_{j\pm}(\varphi) &= - |\Delta| \cos(\frac{\varphi}{2} - {s_{j\pm}^{r}}\theta_B  )    \; ,\\
                 \varepsilon^{l}_{j\pm}(\varphi) &= - \varepsilon^{r}_{j\mp}(\varphi)  \; ,
             \end{aligned} 
         \end{equation} 
        such that the amount each Andreev bound state is shifted by is entirely determined by the spin-projection of its constituent linearised electron state, $s_{j\pm}^{r}$. Considering this, the striking observation of Fig. \ref{fig::multiBandABS} is that when the spin-orbit strength is increased such that $k_\alpha \sim k_1$, there appears to be only a single state phase shifted to the right, the rest are phase shifted towards the left. Qualitatively, this agrees with Fig. \ref{fig::RashbaShift}.b where we found that at this spin-orbit coupling strength, almost every occupied right-moving state has a positive spin-projection. As a result, almost every Andreev bound state will be shifted in the same direction under the external magnetic field. Similarly, when the spin-orbit strength is increased further such that $k_\alpha = 1.5k_1$, the normalised density of states appears similar to the system with no spin-orbit coupling. This is again predicted by Fig. \ref{fig::RashbaShift}.c where we found the spin-bands decouple such that both spin-up and spin-down states are again occupied. Due to the occupation of both states with positive and negative spin-projection, there will be Andreev bound states present which are shifted in either direction as observed.
                \begin{figure}[ht]
                    \centering
                    \includegraphics[width=0.5\textwidth]{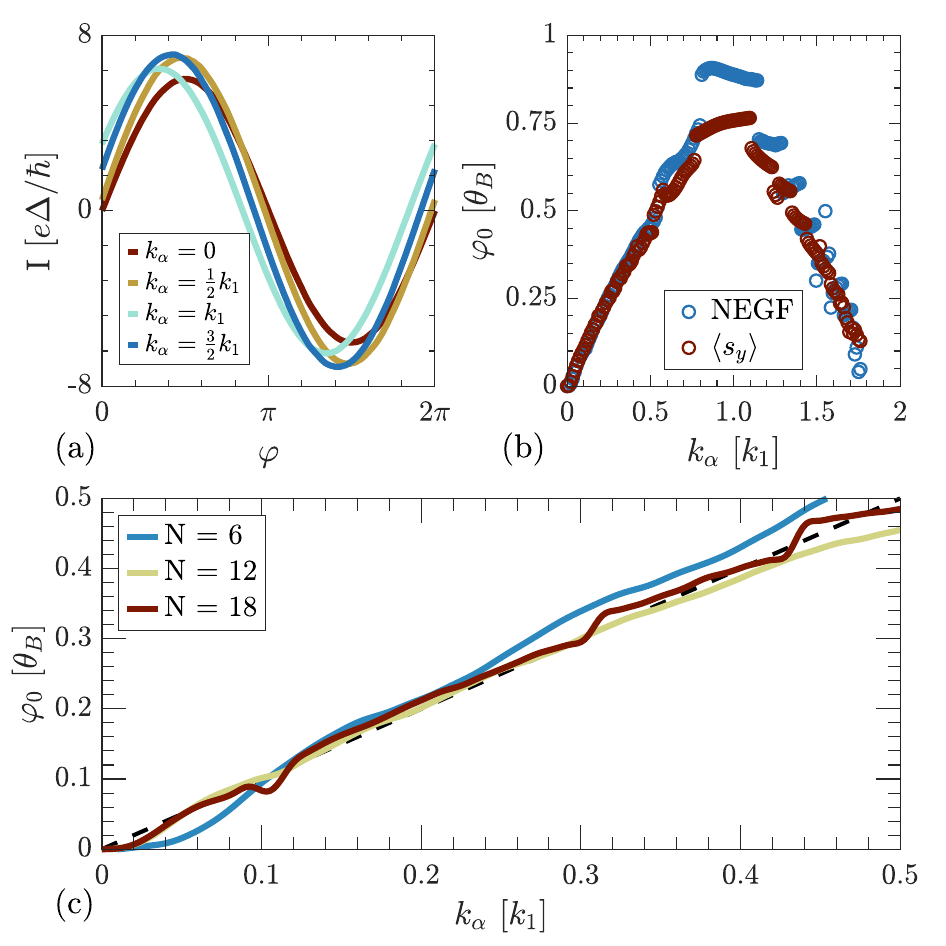} 
                    \caption{(a) Josephson current computed using Eq. \ref{eq::CurrentFromLead} for the system outlined in Sec. \ref{ssec::MultiModeNumerics}. (b) The corresponding anomalous phase extracted from the Josephson current. The blue markers indicate the non-equilibrium Green function (NEGF) calculation, whilst the overlaid red markers indicate the dispersion relation calculation for the average spin-projection outlined in Sec. \ref{ssec::BandStructure2D}. (c) The anomalous phase as a function of the spin-orbit coupling strength computed from the NEGF calculation for three system with different numbers of occupied subbands. To change the number of occupied subbands, every parameter remained the same except the Fermi level which was manually adjusted.}
                    \label{fig::multiBandCurrent}
                \end{figure} 

        Solving for the current given by Eq. \ref{eq::CurrentFromLead}, we plot the Josephson current corresponding to each figure within Fig. \ref{fig::multiBandABS} in Fig. \ref{fig::multiBandCurrent}.a. In Fig. \ref{fig::multiBandCurrent}.a we observe that the sinusoidal Josephson relation experiences a phase shift as the spin-orbit strength is increased, however, at larger values the phase shift appears to decreases. To study this behaviour, we extract the phase shift from the Josephson current and plot them in Fig. \ref{fig::multiBandCurrent}.b. Overlaid on top of the NEGF calculations, we have also plotted the average spin-projection $\left< s_y \right>$ for the same physical system numerically computed using the dispersion relation calculation of Sec. \ref{ssec::BandStructure2D}. It is important to note that although both curves are computational, the NEGF results are solving the dynamics of the system such that we can explicitly extract the anomalous phase, whilst the dispersion relation calculation is simply computing the average spin-projection of the occupied right-moving states. The fact that they are similar is strong evidence that the spin-texture is the predominant effect in the anomalous Josephson effect for ballistic systems.

        An important feature of the predicted anomalous phase was that for spin orbit strengths where $k_{\alpha} < k_1$, the anomalous phase is predicted to be linear in the spin-orbit strength. To check this, in Fig. \ref{fig::multiBandCurrent}.c we plot the anomalous phase extracted from the NEGF calculations within this regime for systems with different number of occupied states. Overall, we observe that the behaviour is highly linear, with the slope appearing to approach unity as the number of subbands increases; effectively identical to the prediction made in Fig. \ref{fig::SpinExpect}.b.
        
        Considering the other predictions made in Sec. \ref{ssec::AnomalousPhaseMulti}, for spin-orbit strengths where $k_{\alpha} \sim k_1$ we observe in Fig. \ref{fig::multiBandCurrent}.b a rounding of the anomalous phase such that it reaches a plateau. In this regime the anomalous phase is effectively independent of the value of the spin-orbit strength $\varphi \sim \theta_B$ as predicted. Similarly, for larger spin-orbit strengths we observe that the numerical anomalous phase decays linearly before decaying to zero as predicted. It is important to note that these qualitative results were highly robust to variations in the system such as varying the number $N$ of occupied states or changing lattice parameters such as the lattice spacing $a$ or effective mass $m^*$. However, the exact nature of the anomalous phase, most particularly the discrete jumps observed as the spin-orbit coupling strength was increased -- due to the number of occupied states within the system changing -- did differ slightly between computational samples.



    %
    
%

%
\section{Conclusions and discussion}\label{sec::Conclusion}

    In this work, we have studied the anomalous Josephson effect in quasi one-dimensional ballistic structures with both two-dimensional Rashba spin-orbit coupling and an external magnetic field. We constructed a microscopic model for the resulting dynamics by focusing solely on the role of the spin-texture which forms due to the coupling between adjacent transverse subbands of opposite spin. This microscopic model found great agreement when compared to computational non-equilibrium Green function calculations. We note our results are only applicable for a subset of experimental devices; we are working purely in the ballistic regime, are studying only the short junction limit, and have ignored the vector potential. The purpose of this was to focus solely on the role of spin-texture asymmetry, which represents an important area of study within the field of proximitised nanowire devices. 

    For the case of systems with only a single occupied subband we analytically derived a closed-form expression -- albeit requiring the analytic solution of a quartic polynomial -- for the anomalous phase produced. This formula was very similar to previous results in the literature, however, it explicitly included a term which factors in the role of spin mixing. Indeed, we found that the asymmetry in this spin-mixing is the largest driver of the anomalous phase, rather than asymmetry in the Fermi velocities as has been the main subject of discussion in the literature. 

    For the case of systems with multiple occupied subbands, we explicitly ignored the Fermi velocity asymmetry to solely focus on the role of the spin-texture within the system. Based on this microscopic model, we can list four theoretical predictions:
        \begin{enumerate}
            \item The anomalous phase should be linear in the external magnetic field for Zeeman energies weaker than the subband splitting. 
            \item For weaker spin-orbit strengths, $\alpha < \hbar v_F$, the anomalous phase should be linear in the spin-orbit strength. 
            \item For an appropriately large spin-orbit strength ($\alpha \sim \hbar v_F$), and under a constant magnetic field, the anomalous phase should parabolically reach a maximum value. 
            \item In the limit of large spin-orbit strengths, the anomalous phase is predicted to decay to zero.
    
        \end{enumerate} 
    Each of these predictions were directly observed in numerical calculations computing using the non-equilibrium Green function formalism. As a result, it appears that for ballistic systems the spin-texture induced by the Rashba spin-orbit coupling is the critical microscopic quantity which produces the anomalous Josephson effect. 
    
    Interestingly, there has been experimental evidence for the first three theoretical predictions \cite{mayerGateControlledAnomalous2020,strambiniJosephsonPhaseBattery2020,reinhardtLinkSupercurrentDiode2023}. Perhaps most striking is our prediction for the anomalous phase to briefly saturate as a function of the spin-orbit strength when $\alpha \sim \hbar v_F$, where similar behaviour was observed in Ref. \cite{mayerGateControlledAnomalous2020}. In that work, they noted that the origin of that observation was unknown. On the possibility of experimentally observing the final prediction -- namely, that the anomalous phase should vanish in the large spin-orbit coupling limit -- we note that for InAs systems that it is experimentally possible to tune the spin-orbit coupling strength within two orders of magnitude hence this limit may be observable for certain systems \cite{wickramasingheTransportPropertiesSurface2018}.

    Recently the anomalous Josephson effect has been experimentally linked to the Josephson diode effect \cite{reinhardtLinkSupercurrentDiode2023}. Hence in future work we look to explore explicitly this link between the spin-texture induced by Rashba spin-orbit coupling and critical current asymmetry.  
\section*{Acknowledgments}\label{sec::Ack}
    We acknowledge helpful discussions with both Francesco Giazotto and Elia Strambini concerning the anomalous Josephson effect. We also acknowledge discussions with Tyler Whittaker on the formation of spin-textures in semi-conducting nanowires.

\appendix

\section{Andreev bound state formation}\label{sec::ABSformation}
    
    Considering the Josephson junction of length $L$ illustrated in Fig. \ref{fig::roughDiagram}, the interface between the three regions occur at $\pm L / 2$. Focusing on the non-superconducting central region, after factoring out the fast oscillations on the order of $k_j$, the linearised eigenenergies of Eq. \ref{eq::CentralHam} have the form 
        \begin{equation}\label{eq::EigenEnergiesLinearised}
            \varepsilon(k)  = \hbar v_{j\pm}^{l/r}k  -  s_{j\pm}^{l/r}E_{Z}  \; ,
        \end{equation} 
    We also introduce a spinor $\chi^{l/r}_{j\pm}$ defined such that
    \begin{equation}\label{eq::SpinorDef}
        {\left(\chi^{l/r}_{j\pm}\right)}^{\dagger} \sigma_{y}{\chi^{l/r}_{j\pm}} = {s^{l/r}_{j\pm}}  \; ,
    \end{equation} 
   As a result, the travelling wave solution for a left or right moving electron with energy $\varepsilon$ in the normal region is given by 
    \begin{equation}\label{eq::TravellingElectron}
        \ket{\psi^{l/r}_{j\pm}(x)}_\text{electron} = \chi^{l/r}_{j\pm} e^{i\left[(\varepsilon+s_{j\pm}^{l/r}E_{Z}) / \hbar v_{j\pm}^{l/r}\right] x}  \; ,
    \end{equation} 
   the hole states are then given by
   \begin{equation}\label{eq::TravellingElectron}
    \ket{\psi^{l/r}_{j\pm}(x)}_\text{hole} = i\sigma_{y}\chi^{l/r}_{j\pm} e^{-i\left[(\varepsilon+s_{j\pm}^{l/r}E_{Z}) / \hbar v_{j\pm}^{l/r}\right] x}  \; ,
    \end{equation} 

   The same can be done for the proximitised superconducting regions by evaluating the eigenstates of Eq. \ref{eq::LeftRightFull}. However, as we are interested in energies within the superconducting gap, $\varepsilon<|\Delta|$, then the only allowed states are evanescent states which decay into the superconductors. The resulting eigenstates in the left (L) and right (R) superconductors are given by \cite{saulsAndreevBoundStates2018}
    \begin{equation}\label{eq::SuperconductinStates}
        \begin{aligned}
            \ket{ \psi^{l/r}_{j\pm}(x) }_L& =     
            \begin{pmatrix}
            (\varepsilon - i\Lambda)\chi^{l/r}_{j\pm}\\\Delta_L ^{*}( i\sigma_{y})\chi^{l/r}_{j\pm}
            \end{pmatrix}
            e^{-\Lambda (x+L / 2)/\hbar v^{l/r}_{j\pm}}   \; ,  \\
            \ket{ \psi^{l/r}_{j\pm}(x) }_R& =     
            \begin{pmatrix}
            (\varepsilon - i\Lambda)\chi^{l/r}_{j\pm}\\\Delta_R ^{*}( i\sigma_{y})\chi^{l/r}_{j\pm}
            \end{pmatrix}
            e^{-\Lambda (x-L / 2)/\hbar v^{l/r}_{j\pm}}   \; , 
        \end{aligned} 
    \end{equation} 
   where 
    \begin{equation}\label{eq::Lambda}
        \Lambda \equiv \sqrt{|\Delta|^2 - \varepsilon^2}  \; .
    \end{equation}
   and we note that we are now working in a particle-hole basis. Within this basis, the total wavefunction for the non-superconducting central (C) region, consisting of counter-propagating electrons and holes, is given by the superposition
    \begin{equation}\label{eq::TravellingElectron}
            \begin{aligned}
                \ket{\psi^{l/r}_{j\pm}(x)}_\text{C} = A\begin{pmatrix}
                    \chi^{l/r}_{j\pm}\\0
                     \end{pmatrix}e^{i\left[(\varepsilon+s_{j\pm}^{l/r}E_{Z}) / \hbar v_{j\pm}^{l/r}\right] x}\\
                    + B\begin{pmatrix}
                        0\\i\sigma_y\chi^{l/r}_{j\pm}
                         \end{pmatrix}e^{-i\left[(\varepsilon+s_{j\pm}^{l/r}E_{Z}) / \hbar v_{j\pm}^{l/r}\right] x} \; .
            \end{aligned} 
    \end{equation}

   By matching the wavefunctions at $x = \pm L / 2$ -- a necessary condition for the formation of a bound state -- we find that
        \begin{equation}\label{eq::MatchedWave}
            \frac{\varepsilon-i\Lambda}{\varepsilon+i\Lambda} = \frac{\Delta_{L}}{\Delta_{R}}e^{-2iL\varepsilon / \hbar v_{j\pm}^{l/r}}e^{-2iLs_{j\pm}^{l/r}E_{Z} / \hbar v_{j\pm}^{l/r}}  \; .
        \end{equation} 
    Utilising some identities, this can be rewritten as
        \begin{equation}\label{eq::TrigIdentityMatched}
            e^{-2i\text{acos}(\varepsilon / |\Delta|)} = e^{i\varphi} e^{-2iL\varepsilon / \hbar v_{j\pm}^{l/r}}e^{-2iLs_{j\pm}^{l/r}E_{Z} /\hbar v_{j\pm}^{l/r}} \; ,
        \end{equation} 
    as in the maintext.

\section{Linear average spin-projection}\label{sec::LinearAverageSpin}

\begin{figure}[ht]
        \centering
        \includegraphics[width=0.5\textwidth]{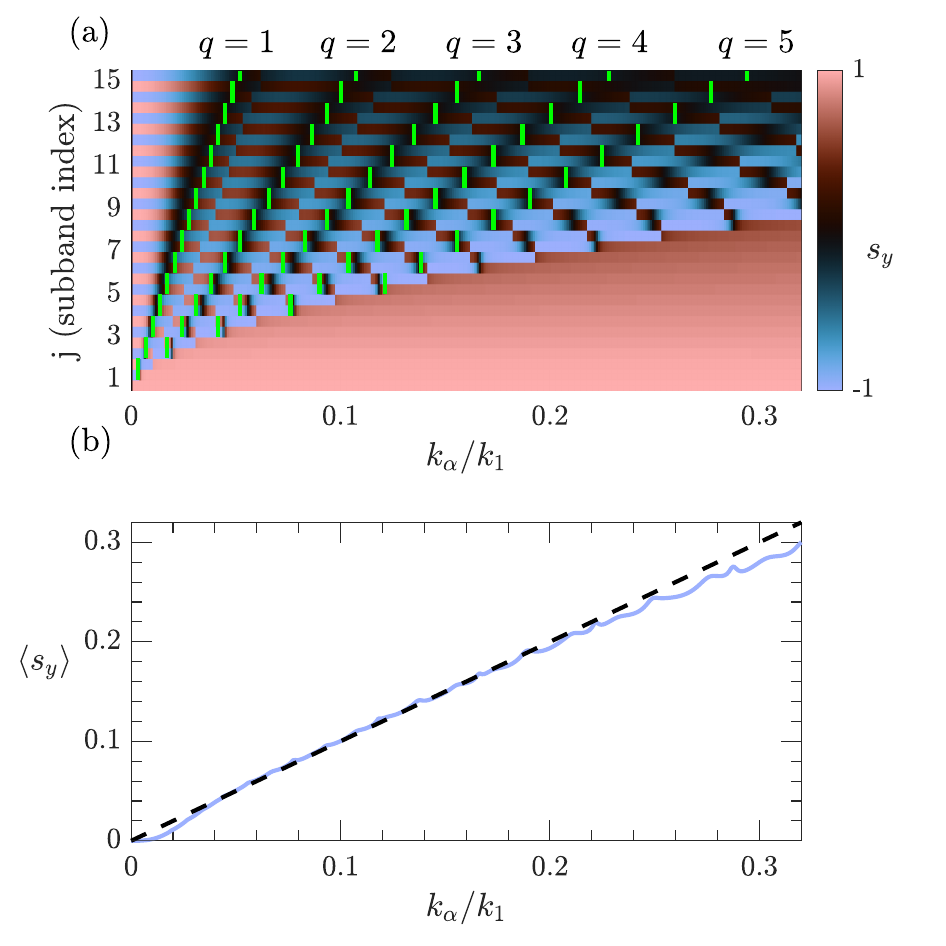}
        \caption{(a) The spin-projection of each state $s_j$ plotted as a function of the spin-orbit coupling strength. The black regions indicate where a spin-projection has changed signs. Overlaid in green is the critical spin-orbit strength -- given by Eq. \ref{eq::FermiWavevectorApproxLinear} -- where we naïvely expect a sign flip to occur. (b) The average spin-projection of the system defined in (a).}
        \label{fig::spinProj}
    \end{figure}

To understand why the average spin-projection is approximately linear in the spin-orbit coupling strength, we first note that what perturbs the average spin-projection is not the mixing between occupied states -- as any mixing does would not change the total sum of the occupied spin-projections -- but rather unoccupied higher subbands which mix with an occupied state. Hence the critical parameter to determine is at what spin-orbit coupling strengths does some unoccupied state couple with an occupied state. To determine at what specific spin-orbit value a higher subband mixes with an occupied state, it is useful to briefly return to the case of a one-dimensional Rashba spin-orbit coupling where there is no overlap between adjacent subbands. In this case, for a system with $N$ states, the Fermi-wavevectors of each subband can be computed analytically and are given by
    \begin{equation}\label{eq::FermiWavevectorShiftApprox}
        k_{j\pm}^{r} =  \frac{\pi}{W}\sqrt{ N^{2} - (j-1)^{2} }\pm k_{\alpha} \; ,
    \end{equation} 
As a result, the $j^{\text{th}}$ spin-up state overlaps with the $(j+q)^{\text{th}}$ spin-down at the Fermi level when the spin-orbit strength is tuned such that
    \begin{equation}\label{eq::FermiWavevectorApprox}
        k_\alpha  = \frac{\pi}{2W}\left[ \sqrt{ N^2 -  (j-1)^{2} }  - \sqrt{ N^2 - (j+q-1)^2  } \right]  \; ,
    \end{equation} 
Pulling out a factor or $k_1 = N \pi / W$, and Taylor expanding the roots, we can write the Rashba wave-vector required to overlap the states at the Fermi level as
\begin{equation}\label{eq::FermiWavevectorApproxLinear}
    \frac{	k_{\alpha}}{k_{1}}  \sim \frac{qj}{2N^2}      
\end{equation} 
Returning to the proper two-dimensional Rashba interaction -- such that overlapping bands form avoided crossings -- the spin-orbit strengths defined in Eq. \ref{eq::FermiWavevectorApproxLinear} now define critical values where we expect to observe a spin-flip between the $j^{\text{th}}$ state and its $q^{\text{th}}$ neighbour. To demonstrate this, we plot the different spin-projections of each occupied right-moving state as the spin-orbit coupling strength is increased a system with $N=15$ occupied subbands (thirty occupied spin states) in Fig. \ref{fig::spinProj}.a. Overlaid on top is the critical spin-orbit strength defined in Eq. \ref{eq::FermiWavevectorApproxLinear}. Visually it provides a good approximation to when the spin-projections flip sign.
Unoccupied higher subband states always couple initially with the highest occupied state -- namely, the $j=N$ state. As a result, using Eq. \ref{eq::FermiWavevectorApproxLinear} the highest occupied subband will be fully mixed with the $N+q$ subband when $k_\alpha \sim qk_1 / 2N$. Although this mixing likely obeys a complicated sigmoid as in Eq. \ref{eq::SpinAndVel} for the single subband case, we can consider an ideal system where the states begin mixing initially when $k_\alpha = 0$, are fully mixed when $k_\alpha \sim qk_1 / 2N$, and doubling this, have experienced a full sign change in the spin-projection when $k_\alpha \sim qk_1 / N$. As a result, as the Rashba wavevector $k_\alpha$ is increased, every multiple of ${k_1}/{N}$ the total spin of the right-moving occupied states increases by two. We can approximate these jumps by the use of the Heaviside step function such that
    \begin{equation}\label{eq::AverageAsSum}
      \left< s_y \right> \sim \frac{1}{2N}\sum_{j=1}^N 2\Theta\left( \frac{k_{\alpha}}{k_{1}} - \frac{j}{N} \right)   \; ,
    \end{equation} 
In the limit as $N\to \infty$ we can replace the discrete sum with an integral -- note, we have replaced the discrete $j / N$ with the continues variable $z$
\begin{equation}\label{eq::AverageAsSumIntegral}
        \begin{aligned}
            \left< s_y \right> & \underset{N \to \infty}{\longrightarrow} \frac{1}{2}\int_{0}^1   2\Theta\left( \frac{k_{\alpha}}{k_{1}} - z \right) \; dz  \; , \\ 
            &= \frac{k_\alpha}{k_1} \; .
        \end{aligned} 
\end{equation} 
To illustrate the effectiveness of this approximation, in Fig. \ref{fig::spinProj}.b we plot the average spin-projection for the $N=15$ system shown in Fig. \ref{fig::spinProj}.a. We visually observe a very good agreement with the linear approximation. 
\bibliography{/Users/ross.monaghan/Documents/System/Resources/Bib/references.bib}

\begin{thebibliography}{44}%
\makeatletter
\providecommand \@ifxundefined [1]{%
 \@ifx{#1\undefined}
}%
\providecommand \@ifnum [1]{%
 \ifnum #1\expandafter \@firstoftwo
 \else \expandafter \@secondoftwo
 \fi
}%
\providecommand \@ifx [1]{%
 \ifx #1\expandafter \@firstoftwo
 \else \expandafter \@secondoftwo
 \fi
}%
\providecommand \natexlab [1]{#1}%
\providecommand \enquote  [1]{``#1''}%
\providecommand \bibnamefont  [1]{#1}%
\providecommand \bibfnamefont [1]{#1}%
\providecommand \citenamefont [1]{#1}%
\providecommand \href@noop [0]{\@secondoftwo}%
\providecommand \href [0]{\begingroup \@sanitize@url \@href}%
\providecommand \@href[1]{\@@startlink{#1}\@@href}%
\providecommand \@@href[1]{\endgroup#1\@@endlink}%
\providecommand \@sanitize@url [0]{\catcode `\\12\catcode `\$12\catcode
  `\&12\catcode `\#12\catcode `\^12\catcode `\_12\catcode `\%12\relax}%
\providecommand \@@startlink[1]{}%
\providecommand \@@endlink[0]{}%
\providecommand \url  [0]{\begingroup\@sanitize@url \@url }%
\providecommand \@url [1]{\endgroup\@href {#1}{\urlprefix }}%
\providecommand \urlprefix  [0]{URL }%
\providecommand \Eprint [0]{\href }%
\providecommand \doibase [0]{https://doi.org/}%
\providecommand \selectlanguage [0]{\@gobble}%
\providecommand \bibinfo  [0]{\@secondoftwo}%
\providecommand \bibfield  [0]{\@secondoftwo}%
\providecommand \translation [1]{[#1]}%
\providecommand \BibitemOpen [0]{}%
\providecommand \bibitemStop [0]{}%
\providecommand \bibitemNoStop [0]{.\EOS\space}%
\providecommand \EOS [0]{\spacefactor3000\relax}%
\providecommand \BibitemShut  [1]{\csname bibitem#1\endcsname}%
\let\auto@bib@innerbib\@empty
\bibitem [{\citenamefont {Josephson}(1962)}]{josephsonPossibleNewEffects1962}%
  \BibitemOpen
  \bibfield  {author} {\bibinfo {author} {\bibfnamefont {B.~D.}\ \bibnamefont
  {Josephson}},\ }\bibfield  {title} {\bibinfo {title} {Possible new effects in
  superconductive tunnelling},\ }\href
  {https://doi.org/10.1016/0031-9163(62)91369-0} {\bibfield  {journal}
  {\bibinfo  {journal} {Physics Letters}\ }\textbf {\bibinfo {volume} {1}},\
  \bibinfo {pages} {251} (\bibinfo {year} {1962})}\BibitemShut {NoStop}%
\bibitem [{\citenamefont {Bulaevskii}\ \emph {et~al.}(1977)\citenamefont
  {Bulaevskii}, \citenamefont {Kuzii},\ and\ \citenamefont
  {Sobyanin}}]{bulaevskiiSuperconductingSystemWeak1977}%
  \BibitemOpen
  \bibfield  {author} {\bibinfo {author} {\bibfnamefont {L.~N.}\ \bibnamefont
  {Bulaevskii}}, \bibinfo {author} {\bibfnamefont {V.~V.}\ \bibnamefont
  {Kuzii}},\ and\ \bibinfo {author} {\bibfnamefont {A.~A.}\ \bibnamefont
  {Sobyanin}},\ }\bibfield  {title} {\bibinfo {title} {Superconducting system
  with weak coupling to the current in the ground state},\ }\href
  {https://www.osti.gov/biblio/7316063} {\bibfield  {journal} {\bibinfo
  {journal} {JETP Lett. (USSR) (Engl. Transl.); (United States)}\ }\textbf
  {\bibinfo {volume} {25:7}} (\bibinfo {year} {1977})}\BibitemShut {NoStop}%
\bibitem [{\citenamefont {Rasmussen}\ \emph {et~al.}(2016)\citenamefont
  {Rasmussen}, \citenamefont {Danon}, \citenamefont {Suominen}, \citenamefont
  {Nichele}, \citenamefont {Kjaergaard},\ and\ \citenamefont
  {Flensberg}}]{rasmussenEffectsSpinorbitCoupling2016}%
  \BibitemOpen
  \bibfield  {author} {\bibinfo {author} {\bibfnamefont {A.}~\bibnamefont
  {Rasmussen}}, \bibinfo {author} {\bibfnamefont {J.}~\bibnamefont {Danon}},
  \bibinfo {author} {\bibfnamefont {H.}~\bibnamefont {Suominen}}, \bibinfo
  {author} {\bibfnamefont {F.}~\bibnamefont {Nichele}}, \bibinfo {author}
  {\bibfnamefont {M.}~\bibnamefont {Kjaergaard}},\ and\ \bibinfo {author}
  {\bibfnamefont {K.}~\bibnamefont {Flensberg}},\ }\bibfield  {title} {\bibinfo
  {title} {Effects of spin-orbit coupling and spatial symmetries on the
  {{Josephson}} current in {{SNS}} junctions},\ }\href
  {https://doi.org/10.1103/PhysRevB.93.155406} {\bibfield  {journal} {\bibinfo
  {journal} {Physical Review B}\ }\textbf {\bibinfo {volume} {93}},\ \bibinfo
  {pages} {155406} (\bibinfo {year} {2016})}\BibitemShut {NoStop}%
\bibitem [{\citenamefont {Assouline}\ \emph {et~al.}(2019)\citenamefont
  {Assouline}, \citenamefont {{Feuillet-Palma}}, \citenamefont {Bergeal},
  \citenamefont {Zhang}, \citenamefont {Mottaghizadeh}, \citenamefont
  {Zimmers}, \citenamefont {Lhuillier}, \citenamefont {Eddrie}, \citenamefont
  {Atkinson}, \citenamefont {Aprili},\ and\ \citenamefont
  {Aubin}}]{assoulineSpinOrbitInducedPhaseshift2019}%
  \BibitemOpen
  \bibfield  {author} {\bibinfo {author} {\bibfnamefont {A.}~\bibnamefont
  {Assouline}}, \bibinfo {author} {\bibfnamefont {C.}~\bibnamefont
  {{Feuillet-Palma}}}, \bibinfo {author} {\bibfnamefont {N.}~\bibnamefont
  {Bergeal}}, \bibinfo {author} {\bibfnamefont {T.}~\bibnamefont {Zhang}},
  \bibinfo {author} {\bibfnamefont {A.}~\bibnamefont {Mottaghizadeh}}, \bibinfo
  {author} {\bibfnamefont {A.}~\bibnamefont {Zimmers}}, \bibinfo {author}
  {\bibfnamefont {E.}~\bibnamefont {Lhuillier}}, \bibinfo {author}
  {\bibfnamefont {M.}~\bibnamefont {Eddrie}}, \bibinfo {author} {\bibfnamefont
  {P.}~\bibnamefont {Atkinson}}, \bibinfo {author} {\bibfnamefont
  {M.}~\bibnamefont {Aprili}},\ and\ \bibinfo {author} {\bibfnamefont
  {H.}~\bibnamefont {Aubin}},\ }\bibfield  {title} {\bibinfo {title}
  {Spin-{{Orbit}} induced phase-shift in {{Bi2Se3 Josephson}} junctions},\
  }\href {https://doi.org/10.1038/s41467-018-08022-y} {\bibfield  {journal}
  {\bibinfo  {journal} {Nature Communications}\ }\textbf {\bibinfo {volume}
  {10}},\ \bibinfo {pages} {126} (\bibinfo {year} {2019})}\BibitemShut
  {NoStop}%
\bibitem [{\citenamefont {Strambini}\ \emph {et~al.}(2020)\citenamefont
  {Strambini}, \citenamefont {Iorio}, \citenamefont {Durante}, \citenamefont
  {Citro}, \citenamefont {{Sanz-Fern{\'a}ndez}}, \citenamefont {Guarcello},
  \citenamefont {Tokatly}, \citenamefont {Braggio}, \citenamefont {Rocci},
  \citenamefont {Ligato}, \citenamefont {Zannier}, \citenamefont {Sorba},
  \citenamefont {Bergeret},\ and\ \citenamefont
  {Giazotto}}]{strambiniJosephsonPhaseBattery2020}%
  \BibitemOpen
  \bibfield  {author} {\bibinfo {author} {\bibfnamefont {E.}~\bibnamefont
  {Strambini}}, \bibinfo {author} {\bibfnamefont {A.}~\bibnamefont {Iorio}},
  \bibinfo {author} {\bibfnamefont {O.}~\bibnamefont {Durante}}, \bibinfo
  {author} {\bibfnamefont {R.}~\bibnamefont {Citro}}, \bibinfo {author}
  {\bibfnamefont {C.}~\bibnamefont {{Sanz-Fern{\'a}ndez}}}, \bibinfo {author}
  {\bibfnamefont {C.}~\bibnamefont {Guarcello}}, \bibinfo {author}
  {\bibfnamefont {I.~V.}\ \bibnamefont {Tokatly}}, \bibinfo {author}
  {\bibfnamefont {A.}~\bibnamefont {Braggio}}, \bibinfo {author} {\bibfnamefont
  {M.}~\bibnamefont {Rocci}}, \bibinfo {author} {\bibfnamefont
  {N.}~\bibnamefont {Ligato}}, \bibinfo {author} {\bibfnamefont
  {V.}~\bibnamefont {Zannier}}, \bibinfo {author} {\bibfnamefont
  {L.}~\bibnamefont {Sorba}}, \bibinfo {author} {\bibfnamefont {F.~S.}\
  \bibnamefont {Bergeret}},\ and\ \bibinfo {author} {\bibfnamefont
  {F.}~\bibnamefont {Giazotto}},\ }\bibfield  {title} {\bibinfo {title} {A
  {{Josephson}} phase battery},\ }\href
  {https://doi.org/10.1038/s41565-020-0712-7} {\bibfield  {journal} {\bibinfo
  {journal} {Nature Nanotechnology}\ }\textbf {\bibinfo {volume} {15}},\
  \bibinfo {pages} {656} (\bibinfo {year} {2020})}\BibitemShut {NoStop}%
\bibitem [{\citenamefont {Reinhardt}\ \emph {et~al.}(2023)\citenamefont
  {Reinhardt}, \citenamefont {Ascherl}, \citenamefont {Costa}, \citenamefont
  {Berger}, \citenamefont {Gronin}, \citenamefont {Gardner}, \citenamefont
  {Lindemann}, \citenamefont {Manfra}, \citenamefont {Fabian}, \citenamefont
  {Kochan}, \citenamefont {Strunk},\ and\ \citenamefont
  {Paradiso}}]{reinhardtLinkSupercurrentDiode2023}%
  \BibitemOpen
  \bibfield  {author} {\bibinfo {author} {\bibfnamefont {S.}~\bibnamefont
  {Reinhardt}}, \bibinfo {author} {\bibfnamefont {T.}~\bibnamefont {Ascherl}},
  \bibinfo {author} {\bibfnamefont {A.}~\bibnamefont {Costa}}, \bibinfo
  {author} {\bibfnamefont {J.}~\bibnamefont {Berger}}, \bibinfo {author}
  {\bibfnamefont {S.}~\bibnamefont {Gronin}}, \bibinfo {author} {\bibfnamefont
  {G.~C.}\ \bibnamefont {Gardner}}, \bibinfo {author} {\bibfnamefont
  {T.}~\bibnamefont {Lindemann}}, \bibinfo {author} {\bibfnamefont {M.~J.}\
  \bibnamefont {Manfra}}, \bibinfo {author} {\bibfnamefont {J.}~\bibnamefont
  {Fabian}}, \bibinfo {author} {\bibfnamefont {D.}~\bibnamefont {Kochan}},
  \bibinfo {author} {\bibfnamefont {C.}~\bibnamefont {Strunk}},\ and\ \bibinfo
  {author} {\bibfnamefont {N.}~\bibnamefont {Paradiso}},\ }\href
  {https://doi.org/10.48550/arXiv.2308.01061} {\bibinfo {title} {Link between
  supercurrent diode and anomalous {{Josephson}} effect revealed by
  gate-controlled interferometry}} (\bibinfo {year} {2023}),\ \Eprint
  {https://arxiv.org/abs/2308.01061} {arxiv:2308.01061 [cond-mat]} \BibitemShut
  {NoStop}%
\bibitem [{\citenamefont {Mayer}\ \emph {et~al.}(2020)\citenamefont {Mayer},
  \citenamefont {Dartiailh}, \citenamefont {Yuan}, \citenamefont
  {Wickramasinghe}, \citenamefont {Rossi},\ and\ \citenamefont
  {Shabani}}]{mayerGateControlledAnomalous2020}%
  \BibitemOpen
  \bibfield  {author} {\bibinfo {author} {\bibfnamefont {W.}~\bibnamefont
  {Mayer}}, \bibinfo {author} {\bibfnamefont {M.~C.}\ \bibnamefont
  {Dartiailh}}, \bibinfo {author} {\bibfnamefont {J.}~\bibnamefont {Yuan}},
  \bibinfo {author} {\bibfnamefont {K.~S.}\ \bibnamefont {Wickramasinghe}},
  \bibinfo {author} {\bibfnamefont {E.}~\bibnamefont {Rossi}},\ and\ \bibinfo
  {author} {\bibfnamefont {J.}~\bibnamefont {Shabani}},\ }\bibfield  {title}
  {\bibinfo {title} {Gate controlled anomalous phase shift in {{Al}}/{{InAs
  Josephson}} junctions},\ }\href {https://doi.org/10.1038/s41467-019-14094-1}
  {\bibfield  {journal} {\bibinfo  {journal} {Nature Communications}\ }\textbf
  {\bibinfo {volume} {11}},\ \bibinfo {pages} {212} (\bibinfo {year}
  {2020})}\BibitemShut {NoStop}%
\bibitem [{\citenamefont
  {Shukrinov}(2022)}]{shukrinovAnomalousJosephsonEffect2022}%
  \BibitemOpen
  \bibfield  {author} {\bibinfo {author} {\bibfnamefont {Y.~M.}\ \bibnamefont
  {Shukrinov}},\ }\bibfield  {title} {\bibinfo {title} {Anomalous {{Josephson}}
  effect},\ }\href {https://doi.org/10.3367/UFNe.2020.11.038894} {\bibfield
  {journal} {\bibinfo  {journal} {Physics-Uspekhi}\ }\textbf {\bibinfo {volume}
  {65}},\ \bibinfo {pages} {317} (\bibinfo {year} {2022})}\BibitemShut
  {NoStop}%
\bibitem [{\citenamefont {Moroz}\ and\ \citenamefont
  {Barnes}(1999)}]{morozEffectSpinorbitInteraction1999}%
  \BibitemOpen
  \bibfield  {author} {\bibinfo {author} {\bibfnamefont {A.~V.}\ \bibnamefont
  {Moroz}}\ and\ \bibinfo {author} {\bibfnamefont {C.~H.~W.}\ \bibnamefont
  {Barnes}},\ }\bibfield  {title} {\bibinfo {title} {Effect of the spin-orbit
  interaction on the band structure and conductance of quasi-one-dimensional
  systems},\ }\href {https://doi.org/10.1103/PhysRevB.60.14272} {\bibfield
  {journal} {\bibinfo  {journal} {Physical Review B}\ }\textbf {\bibinfo
  {volume} {60}},\ \bibinfo {pages} {14272} (\bibinfo {year}
  {1999})}\BibitemShut {NoStop}%
\bibitem [{\citenamefont {Bercioux}\ and\ \citenamefont
  {Lucignano}(2015)}]{berciouxQuantumTransportRashba2015}%
  \BibitemOpen
  \bibfield  {author} {\bibinfo {author} {\bibfnamefont {D.}~\bibnamefont
  {Bercioux}}\ and\ \bibinfo {author} {\bibfnamefont {P.}~\bibnamefont
  {Lucignano}},\ }\bibfield  {title} {\bibinfo {title} {Quantum transport in
  {{Rashba}} spin--orbit materials: A review},\ }\href
  {https://doi.org/10.1088/0034-4885/78/10/106001} {\bibfield  {journal}
  {\bibinfo  {journal} {Reports on Progress in Physics}\ }\textbf {\bibinfo
  {volume} {78}},\ \bibinfo {pages} {106001} (\bibinfo {year}
  {2015})}\BibitemShut {NoStop}%
\bibitem [{\citenamefont {Governale}\ and\ \citenamefont
  {Z{\"u}licke}(2002)}]{governaleSpinAccumulationQuantum2002}%
  \BibitemOpen
  \bibfield  {author} {\bibinfo {author} {\bibfnamefont {M.}~\bibnamefont
  {Governale}}\ and\ \bibinfo {author} {\bibfnamefont {U.}~\bibnamefont
  {Z{\"u}licke}},\ }\bibfield  {title} {\bibinfo {title} {Spin accumulation in
  quantum wires with strong {{Rashba}} spin-orbit coupling},\ }\href
  {https://doi.org/10.1103/PhysRevB.66.073311} {\bibfield  {journal} {\bibinfo
  {journal} {Physical Review B}\ }\textbf {\bibinfo {volume} {66}},\ \bibinfo
  {pages} {073311} (\bibinfo {year} {2002})}\BibitemShut {NoStop}%
\bibitem [{\citenamefont {Murani}\ \emph {et~al.}(2017)\citenamefont {Murani},
  \citenamefont {Chepelianskii}, \citenamefont {Gu{\'e}ron},\ and\
  \citenamefont {Bouchiat}}]{muraniAndreevSpectrumHigh2017}%
  \BibitemOpen
  \bibfield  {author} {\bibinfo {author} {\bibfnamefont {A.}~\bibnamefont
  {Murani}}, \bibinfo {author} {\bibfnamefont {A.}~\bibnamefont
  {Chepelianskii}}, \bibinfo {author} {\bibfnamefont {S.}~\bibnamefont
  {Gu{\'e}ron}},\ and\ \bibinfo {author} {\bibfnamefont {H.}~\bibnamefont
  {Bouchiat}},\ }\bibfield  {title} {\bibinfo {title} {Andreev spectrum with
  high spin-orbit interactions: {{Revealing}} spin splitting and topologically
  protected crossings},\ }\href {https://doi.org/10.1103/PhysRevB.96.165415}
  {\bibfield  {journal} {\bibinfo  {journal} {Physical Review B}\ }\textbf
  {\bibinfo {volume} {96}},\ \bibinfo {pages} {165415} (\bibinfo {year}
  {2017})}\BibitemShut {NoStop}%
\bibitem [{\citenamefont {Park}\ and\ \citenamefont
  {Yeyati}(2017)}]{parkAndreevSpinQubits2017}%
  \BibitemOpen
  \bibfield  {author} {\bibinfo {author} {\bibfnamefont {S.}~\bibnamefont
  {Park}}\ and\ \bibinfo {author} {\bibfnamefont {A.~L.}\ \bibnamefont
  {Yeyati}},\ }\bibfield  {title} {\bibinfo {title} {Andreev spin qubits in
  multichannel {{Rashba}} nanowires},\ }\href
  {https://doi.org/10.1103/PhysRevB.96.125416} {\bibfield  {journal} {\bibinfo
  {journal} {Physical Review B}\ }\textbf {\bibinfo {volume} {96}},\ \bibinfo
  {pages} {125416} (\bibinfo {year} {2017})}\BibitemShut {NoStop}%
\bibitem [{\citenamefont {Reynoso}\ \emph {et~al.}(2008)\citenamefont
  {Reynoso}, \citenamefont {Usaj}, \citenamefont {Balseiro}, \citenamefont
  {Feinberg},\ and\ \citenamefont
  {Avignon}}]{reynosoAnomalousJosephsonCurrent2008}%
  \BibitemOpen
  \bibfield  {author} {\bibinfo {author} {\bibfnamefont {A.~A.}\ \bibnamefont
  {Reynoso}}, \bibinfo {author} {\bibfnamefont {G.}~\bibnamefont {Usaj}},
  \bibinfo {author} {\bibfnamefont {C.~A.}\ \bibnamefont {Balseiro}}, \bibinfo
  {author} {\bibfnamefont {D.}~\bibnamefont {Feinberg}},\ and\ \bibinfo
  {author} {\bibfnamefont {M.}~\bibnamefont {Avignon}},\ }\bibfield  {title}
  {\bibinfo {title} {Anomalous {{Josephson Current}} in {{Junctions}} with
  {{Spin Polarizing Quantum Point Contacts}}},\ }\href
  {https://doi.org/10.1103/PhysRevLett.101.107001} {\bibfield  {journal}
  {\bibinfo  {journal} {Physical Review Letters}\ }\textbf {\bibinfo {volume}
  {101}},\ \bibinfo {pages} {107001} (\bibinfo {year} {2008})}\BibitemShut
  {NoStop}%
\bibitem [{\citenamefont {Reynoso}\ \emph {et~al.}(2012)\citenamefont
  {Reynoso}, \citenamefont {Usaj}, \citenamefont {Balseiro}, \citenamefont
  {Feinberg},\ and\ \citenamefont
  {Avignon}}]{reynosoSpinorbitinducedChiralityAndreev2012}%
  \BibitemOpen
  \bibfield  {author} {\bibinfo {author} {\bibfnamefont {A.~A.}\ \bibnamefont
  {Reynoso}}, \bibinfo {author} {\bibfnamefont {G.}~\bibnamefont {Usaj}},
  \bibinfo {author} {\bibfnamefont {C.~A.}\ \bibnamefont {Balseiro}}, \bibinfo
  {author} {\bibfnamefont {D.}~\bibnamefont {Feinberg}},\ and\ \bibinfo
  {author} {\bibfnamefont {M.}~\bibnamefont {Avignon}},\ }\bibfield  {title}
  {\bibinfo {title} {Spin-orbit-induced chirality of {{Andreev}} states in
  {{Josephson}} junctions},\ }\href
  {https://doi.org/10.1103/PhysRevB.86.214519} {\bibfield  {journal} {\bibinfo
  {journal} {Physical Review B}\ }\textbf {\bibinfo {volume} {86}},\ \bibinfo
  {pages} {214519} (\bibinfo {year} {2012})}\BibitemShut {NoStop}%
\bibitem [{\citenamefont {Tosi}\ \emph {et~al.}(2019)\citenamefont {Tosi},
  \citenamefont {Metzger}, \citenamefont {Goffman}, \citenamefont {Urbina},
  \citenamefont {Pothier}, \citenamefont {Park}, \citenamefont {Yeyati},
  \citenamefont {Nyg{\aa}rd},\ and\ \citenamefont
  {Krogstrup}}]{tosiSpinOrbitSplittingAndreev2019}%
  \BibitemOpen
  \bibfield  {author} {\bibinfo {author} {\bibfnamefont {L.}~\bibnamefont
  {Tosi}}, \bibinfo {author} {\bibfnamefont {C.}~\bibnamefont {Metzger}},
  \bibinfo {author} {\bibfnamefont {M.~F.}\ \bibnamefont {Goffman}}, \bibinfo
  {author} {\bibfnamefont {C.}~\bibnamefont {Urbina}}, \bibinfo {author}
  {\bibfnamefont {H.}~\bibnamefont {Pothier}}, \bibinfo {author} {\bibfnamefont
  {S.}~\bibnamefont {Park}}, \bibinfo {author} {\bibfnamefont {A.~L.}\
  \bibnamefont {Yeyati}}, \bibinfo {author} {\bibfnamefont {J.}~\bibnamefont
  {Nyg{\aa}rd}},\ and\ \bibinfo {author} {\bibfnamefont {P.}~\bibnamefont
  {Krogstrup}},\ }\bibfield  {title} {\bibinfo {title} {Spin-{{Orbit
  Splitting}} of {{Andreev States Revealed}} by {{Microwave Spectroscopy}}},\
  }\href {https://doi.org/10.1103/PhysRevX.9.011010} {\bibfield  {journal}
  {\bibinfo  {journal} {Physical Review X}\ }\textbf {\bibinfo {volume} {9}},\
  \bibinfo {pages} {011010} (\bibinfo {year} {2019})}\BibitemShut {NoStop}%
\bibitem [{\citenamefont {Krive}\ \emph {et~al.}(2004)\citenamefont {Krive},
  \citenamefont {Gorelik}, \citenamefont {Shekhter},\ and\ \citenamefont
  {Jonson}}]{kriveChiralSymmetryBreaking2004}%
  \BibitemOpen
  \bibfield  {author} {\bibinfo {author} {\bibfnamefont {I.~V.}\ \bibnamefont
  {Krive}}, \bibinfo {author} {\bibfnamefont {L.~Y.}\ \bibnamefont {Gorelik}},
  \bibinfo {author} {\bibfnamefont {R.~I.}\ \bibnamefont {Shekhter}},\ and\
  \bibinfo {author} {\bibfnamefont {M.}~\bibnamefont {Jonson}},\ }\bibfield
  {title} {\bibinfo {title} {Chiral symmetry breaking and the {{Josephson}}
  current in a ballistic superconductor--quantum wire--superconductor
  junction},\ }\href {https://doi.org/10.1063/1.1739160} {\bibfield  {journal}
  {\bibinfo  {journal} {Low Temperature Physics}\ }\textbf {\bibinfo {volume}
  {30}},\ \bibinfo {pages} {398} (\bibinfo {year} {2004})}\BibitemShut
  {NoStop}%
\bibitem [{\citenamefont {Yokoyama}\ \emph {et~al.}(2014)\citenamefont
  {Yokoyama}, \citenamefont {Eto},\ and\ \citenamefont
  {Nazarov}}]{yokoyamaAnomalousJosephsonEffect2014}%
  \BibitemOpen
  \bibfield  {author} {\bibinfo {author} {\bibfnamefont {T.}~\bibnamefont
  {Yokoyama}}, \bibinfo {author} {\bibfnamefont {M.}~\bibnamefont {Eto}},\ and\
  \bibinfo {author} {\bibfnamefont {Y.~V.}\ \bibnamefont {Nazarov}},\
  }\bibfield  {title} {\bibinfo {title} {Anomalous {{Josephson}} effect induced
  by spin-orbit interaction and {{Zeeman}} effect in semiconductor nanowires},\
  }\href {https://doi.org/10.1103/PhysRevB.89.195407} {\bibfield  {journal}
  {\bibinfo  {journal} {Physical Review B}\ }\textbf {\bibinfo {volume} {89}},\
  \bibinfo {pages} {195407} (\bibinfo {year} {2014})}\BibitemShut {NoStop}%
\bibitem [{\citenamefont {Mironov}\ \emph {et~al.}(2015)\citenamefont
  {Mironov}, \citenamefont {Mel'nikov},\ and\ \citenamefont
  {Buzdin}}]{mironovDoublePathInterference2015}%
  \BibitemOpen
  \bibfield  {author} {\bibinfo {author} {\bibfnamefont {S.~V.}\ \bibnamefont
  {Mironov}}, \bibinfo {author} {\bibfnamefont {A.~S.}\ \bibnamefont
  {Mel'nikov}},\ and\ \bibinfo {author} {\bibfnamefont {A.~I.}\ \bibnamefont
  {Buzdin}},\ }\bibfield  {title} {\bibinfo {title} {Double {{Path
  Interference}} and {{Magnetic Oscillations}} in {{Cooper Pair Transport}}
  through a {{Single Nanowire}}},\ }\href
  {https://doi.org/10.1103/PhysRevLett.114.227001} {\bibfield  {journal}
  {\bibinfo  {journal} {Physical Review Letters}\ }\textbf {\bibinfo {volume}
  {114}},\ \bibinfo {pages} {227001} (\bibinfo {year} {2015})}\BibitemShut
  {NoStop}%
\bibitem [{\citenamefont {Zuo}\ \emph {et~al.}(2017)\citenamefont {Zuo},
  \citenamefont {Mourik}, \citenamefont {Szombati}, \citenamefont {Nijholt},
  \citenamefont {{van Woerkom}}, \citenamefont {Geresdi}, \citenamefont {Chen},
  \citenamefont {Ostroukh}, \citenamefont {Akhmerov}, \citenamefont {Plissard},
  \citenamefont {Car}, \citenamefont {Bakkers}, \citenamefont {Pikulin},
  \citenamefont {Kouwenhoven},\ and\ \citenamefont
  {Frolov}}]{zuoSupercurrentInterferenceFewMode2017}%
  \BibitemOpen
  \bibfield  {author} {\bibinfo {author} {\bibfnamefont {K.}~\bibnamefont
  {Zuo}}, \bibinfo {author} {\bibfnamefont {V.}~\bibnamefont {Mourik}},
  \bibinfo {author} {\bibfnamefont {D.~B.}\ \bibnamefont {Szombati}}, \bibinfo
  {author} {\bibfnamefont {B.}~\bibnamefont {Nijholt}}, \bibinfo {author}
  {\bibfnamefont {D.~J.}\ \bibnamefont {{van Woerkom}}}, \bibinfo {author}
  {\bibfnamefont {A.}~\bibnamefont {Geresdi}}, \bibinfo {author} {\bibfnamefont
  {J.}~\bibnamefont {Chen}}, \bibinfo {author} {\bibfnamefont {V.~P.}\
  \bibnamefont {Ostroukh}}, \bibinfo {author} {\bibfnamefont {A.~R.}\
  \bibnamefont {Akhmerov}}, \bibinfo {author} {\bibfnamefont {S.~R.}\
  \bibnamefont {Plissard}}, \bibinfo {author} {\bibfnamefont {D.}~\bibnamefont
  {Car}}, \bibinfo {author} {\bibfnamefont {E.~P. A.~M.}\ \bibnamefont
  {Bakkers}}, \bibinfo {author} {\bibfnamefont {D.~I.}\ \bibnamefont
  {Pikulin}}, \bibinfo {author} {\bibfnamefont {L.~P.}\ \bibnamefont
  {Kouwenhoven}},\ and\ \bibinfo {author} {\bibfnamefont {S.~M.}\ \bibnamefont
  {Frolov}},\ }\bibfield  {title} {\bibinfo {title} {Supercurrent
  {{Interference}} in {{Few-Mode Nanowire Josephson Junctions}}},\ }\href
  {https://doi.org/10.1103/PhysRevLett.119.187704} {\bibfield  {journal}
  {\bibinfo  {journal} {Physical Review Letters}\ }\textbf {\bibinfo {volume}
  {119}},\ \bibinfo {pages} {187704} (\bibinfo {year} {2017})}\BibitemShut
  {NoStop}%
\bibitem [{\citenamefont
  {De~Gennes}(1964)}]{degennesBoundaryEffectsSuperconductors1964}%
  \BibitemOpen
  \bibfield  {author} {\bibinfo {author} {\bibfnamefont {P.~G.}\ \bibnamefont
  {De~Gennes}},\ }\bibfield  {title} {\bibinfo {title} {Boundary {{Effects}} in
  {{Superconductors}}},\ }\href {https://doi.org/10.1103/RevModPhys.36.225}
  {\bibfield  {journal} {\bibinfo  {journal} {Reviews of Modern Physics}\
  }\textbf {\bibinfo {volume} {36}},\ \bibinfo {pages} {225} (\bibinfo {year}
  {1964})}\BibitemShut {NoStop}%
\bibitem [{\citenamefont
  {Beenakker}(1992)}]{beenakkerThreeUniversalMesoscopic1992}%
  \BibitemOpen
  \bibfield  {author} {\bibinfo {author} {\bibfnamefont {C.~W.~J.}\
  \bibnamefont {Beenakker}},\ }\bibfield  {title} {\bibinfo {title} {Three
  ``{{Universal}}'' {{Mesoscopic Josephson Effects}}},\ }in\ \href
  {https://doi.org/10.1007/978-3-642-84818-6_22} {\emph {\bibinfo {booktitle}
  {Transport {{Phenomena}} in {{Mesoscopic Systems}}}}},\ \bibinfo {series and
  number} {Springer {{Series}} in {{Solid-State Sciences}}},\ \bibinfo {editor}
  {edited by\ \bibinfo {editor} {\bibfnamefont {H.}~\bibnamefont {Fukuyama}}\
  and\ \bibinfo {editor} {\bibfnamefont {T.}~\bibnamefont {Ando}}}\ (\bibinfo
  {publisher} {{Springer}},\ \bibinfo {address} {{Berlin, Heidelberg}},\
  \bibinfo {year} {1992})\ pp.\ \bibinfo {pages} {235--253}\BibitemShut
  {NoStop}%
\bibitem [{\citenamefont {{van Heck}}\ \emph {et~al.}(2017)\citenamefont {{van
  Heck}}, \citenamefont {V{\"a}yrynen},\ and\ \citenamefont
  {Glazman}}]{vanheckZeemanSpinorbitEffects2017}%
  \BibitemOpen
  \bibfield  {author} {\bibinfo {author} {\bibfnamefont {B.}~\bibnamefont {{van
  Heck}}}, \bibinfo {author} {\bibfnamefont {J.~I.}\ \bibnamefont
  {V{\"a}yrynen}},\ and\ \bibinfo {author} {\bibfnamefont {L.~I.}\ \bibnamefont
  {Glazman}},\ }\bibfield  {title} {\bibinfo {title} {Zeeman and spin-orbit
  effects in the {{Andreev}} spectra of nanowire junctions},\ }\href
  {https://doi.org/10.1103/PhysRevB.96.075404} {\bibfield  {journal} {\bibinfo
  {journal} {Physical Review B}\ }\textbf {\bibinfo {volume} {96}},\ \bibinfo
  {pages} {075404} (\bibinfo {year} {2017})}\BibitemShut {NoStop}%
\bibitem [{\citenamefont {Rammer}(2007)}]{rammerQuantumFieldTheory2007}%
  \BibitemOpen
  \bibfield  {author} {\bibinfo {author} {\bibfnamefont {J.}~\bibnamefont
  {Rammer}},\ }\href {https://doi.org/10.1017/CBO9780511618956} {\emph
  {\bibinfo {title} {Quantum {{Field Theory}} of {{Non-equilibrium States}}}}}\
  (\bibinfo  {publisher} {{Cambridge University Press}},\ \bibinfo {address}
  {{Cambridge}},\ \bibinfo {year} {2007})\BibitemShut {NoStop}%
\bibitem [{\citenamefont {Pedder}\ \emph {et~al.}(2016)\citenamefont {Pedder},
  \citenamefont {Meng}, \citenamefont {Tiwari},\ and\ \citenamefont
  {Schmidt}}]{pedderDynamicResponseFunctions2016}%
  \BibitemOpen
  \bibfield  {author} {\bibinfo {author} {\bibfnamefont {C.~J.}\ \bibnamefont
  {Pedder}}, \bibinfo {author} {\bibfnamefont {T.}~\bibnamefont {Meng}},
  \bibinfo {author} {\bibfnamefont {R.~P.}\ \bibnamefont {Tiwari}},\ and\
  \bibinfo {author} {\bibfnamefont {T.~L.}\ \bibnamefont {Schmidt}},\
  }\bibfield  {title} {\bibinfo {title} {Dynamic response functions and helical
  gaps in interacting {{Rashba}} nanowires with and without magnetic fields},\
  }\href {https://doi.org/10.1103/PhysRevB.94.245414} {\bibfield  {journal}
  {\bibinfo  {journal} {Physical Review B}\ }\textbf {\bibinfo {volume} {94}},\
  \bibinfo {pages} {245414} (\bibinfo {year} {2016})}\BibitemShut {NoStop}%
\bibitem [{\citenamefont {Mireles}\ and\ \citenamefont
  {Kirczenow}(2001)}]{mirelesBallisticSpinpolarizedTransport2001}%
  \BibitemOpen
  \bibfield  {author} {\bibinfo {author} {\bibfnamefont {F.}~\bibnamefont
  {Mireles}}\ and\ \bibinfo {author} {\bibfnamefont {G.}~\bibnamefont
  {Kirczenow}},\ }\bibfield  {title} {\bibinfo {title} {Ballistic
  spin-polarized transport and {{Rashba}} spin precession in semiconductor
  nanowires},\ }\href {https://doi.org/10.1103/PhysRevB.64.024426} {\bibfield
  {journal} {\bibinfo  {journal} {Physical Review B}\ }\textbf {\bibinfo
  {volume} {64}},\ \bibinfo {pages} {024426} (\bibinfo {year}
  {2001})}\BibitemShut {NoStop}%
\bibitem [{\citenamefont
  {Andreev}(1964)}]{andreevThermalConductivityIntermediate1964}%
  \BibitemOpen
  \bibfield  {author} {\bibinfo {author} {\bibfnamefont {A.~F.}\ \bibnamefont
  {Andreev}},\ }\bibfield  {title} {\bibinfo {title} {{The thermal conductivity
  of the intermediate state in superconductors}},\ }\href
  {https://www.elibrary.ru/item.asp?id=21796236} {\bibfield  {journal}
  {\bibinfo  {journal} {Zhurnal Eksperimental'noj I Teoreticheskoj Fiziki}\
  }\textbf {\bibinfo {volume} {46}} (\bibinfo {year} {1964})}\BibitemShut
  {NoStop}%
\bibitem [{\citenamefont {Sauls}(2018)}]{saulsAndreevBoundStates2018}%
  \BibitemOpen
  \bibfield  {author} {\bibinfo {author} {\bibfnamefont {J.~A.}\ \bibnamefont
  {Sauls}},\ }\bibfield  {title} {\bibinfo {title} {Andreev bound states and
  their signatures},\ }\href {https://doi.org/10.1098/rsta.2018.0140}
  {\bibfield  {journal} {\bibinfo  {journal} {Philosophical Transactions of the
  Royal Society A: Mathematical, Physical and Engineering Sciences}\ }\textbf
  {\bibinfo {volume} {376}},\ \bibinfo {pages} {20180140} (\bibinfo {year}
  {2018})}\BibitemShut {NoStop}%
\bibitem [{\citenamefont
  {Kulik}(1969)}]{kulikMacroscopicQuantizationProximity1969}%
  \BibitemOpen
  \bibfield  {author} {\bibinfo {author} {\bibfnamefont {I.~O.}\ \bibnamefont
  {Kulik}},\ }\bibfield  {title} {\bibinfo {title} {Macroscopic
  {{Quantization}} and the {{Proximity Effect}} in {{S-N-S Junctions}}},\
  }\href {https://ui.adsabs.harvard.edu/abs/1969JETP...30..944K} {\bibfield
  {journal} {\bibinfo  {journal} {Soviet Journal of Experimental and
  Theoretical Physics}\ }\textbf {\bibinfo {volume} {30}},\ \bibinfo {pages}
  {944} (\bibinfo {year} {1969})}\BibitemShut {NoStop}%
\bibitem [{\citenamefont {Bardeen}\ and\ \citenamefont
  {Johnson}(1972)}]{bardeenJosephsonCurrentFlow1972}%
  \BibitemOpen
  \bibfield  {author} {\bibinfo {author} {\bibfnamefont {J.}~\bibnamefont
  {Bardeen}}\ and\ \bibinfo {author} {\bibfnamefont {J.~L.}\ \bibnamefont
  {Johnson}},\ }\bibfield  {title} {\bibinfo {title} {Josephson {{Current
  Flow}} in {{Pure Superconducting-Normal-Superconducting Junctions}}},\ }\href
  {https://doi.org/10.1103/PhysRevB.5.72} {\bibfield  {journal} {\bibinfo
  {journal} {Physical Review B}\ }\textbf {\bibinfo {volume} {5}},\ \bibinfo
  {pages} {72} (\bibinfo {year} {1972})}\BibitemShut {NoStop}%
\bibitem [{\citenamefont
  {Beenakker}(1991)}]{beenakkerUniversalLimitCriticalcurrent1991}%
  \BibitemOpen
  \bibfield  {author} {\bibinfo {author} {\bibfnamefont {C.~W.~J.}\
  \bibnamefont {Beenakker}},\ }\bibfield  {title} {\bibinfo {title} {Universal
  limit of critical-current fluctuations in mesoscopic {{Josephson}}
  junctions},\ }\href {https://doi.org/10.1103/PhysRevLett.67.3836} {\bibfield
  {journal} {\bibinfo  {journal} {Physical Review Letters}\ }\textbf {\bibinfo
  {volume} {67}},\ \bibinfo {pages} {3836} (\bibinfo {year}
  {1991})}\BibitemShut {NoStop}%
\bibitem [{\citenamefont {Nesterov}\ \emph {et~al.}(2016)\citenamefont
  {Nesterov}, \citenamefont {Houzet},\ and\ \citenamefont
  {Meyer}}]{nesterovAnomalousJosephsonEffect2016}%
  \BibitemOpen
  \bibfield  {author} {\bibinfo {author} {\bibfnamefont {K.~N.}\ \bibnamefont
  {Nesterov}}, \bibinfo {author} {\bibfnamefont {M.}~\bibnamefont {Houzet}},\
  and\ \bibinfo {author} {\bibfnamefont {J.~S.}\ \bibnamefont {Meyer}},\
  }\bibfield  {title} {\bibinfo {title} {Anomalous {{Josephson}} effect in
  semiconducting nanowires as a signature of the topologically nontrivial
  phase},\ }\href {https://doi.org/10.1103/PhysRevB.93.174502} {\bibfield
  {journal} {\bibinfo  {journal} {Physical Review B}\ }\textbf {\bibinfo
  {volume} {93}},\ \bibinfo {pages} {174502} (\bibinfo {year}
  {2016})}\BibitemShut {NoStop}%
\bibitem [{\citenamefont
  {Bagwell}(1992)}]{bagwellSuppressionJosephsonCurrent1992}%
  \BibitemOpen
  \bibfield  {author} {\bibinfo {author} {\bibfnamefont {P.~F.}\ \bibnamefont
  {Bagwell}},\ }\bibfield  {title} {\bibinfo {title} {Suppression of the
  {{Josephson}} current through a narrow, mesoscopic, semiconductor channel by
  a single impurity},\ }\href {https://doi.org/10.1103/PhysRevB.46.12573}
  {\bibfield  {journal} {\bibinfo  {journal} {Physical Review B}\ }\textbf
  {\bibinfo {volume} {46}},\ \bibinfo {pages} {12573} (\bibinfo {year}
  {1992})}\BibitemShut {NoStop}%
\bibitem [{\citenamefont {Bravyi}\ \emph {et~al.}(2011)\citenamefont {Bravyi},
  \citenamefont {DiVincenzo},\ and\ \citenamefont
  {Loss}}]{bravyiSchriefferWolffTransformation2011}%
  \BibitemOpen
  \bibfield  {author} {\bibinfo {author} {\bibfnamefont {S.}~\bibnamefont
  {Bravyi}}, \bibinfo {author} {\bibfnamefont {D.~P.}\ \bibnamefont
  {DiVincenzo}},\ and\ \bibinfo {author} {\bibfnamefont {D.}~\bibnamefont
  {Loss}},\ }\bibfield  {title} {\bibinfo {title} {Schrieffer--{{Wolff}}
  transformation for quantum many-body systems},\ }\href
  {https://doi.org/10.1016/j.aop.2011.06.004} {\bibfield  {journal} {\bibinfo
  {journal} {Annals of Physics}\ }\textbf {\bibinfo {volume} {326}},\ \bibinfo
  {pages} {2793} (\bibinfo {year} {2011})}\BibitemShut {NoStop}%
\bibitem [{\citenamefont {Manchon}\ \emph {et~al.}(2015)\citenamefont
  {Manchon}, \citenamefont {Koo}, \citenamefont {Nitta}, \citenamefont
  {Frolov},\ and\ \citenamefont {Duine}}]{manchonNewPerspectivesRashba2015}%
  \BibitemOpen
  \bibfield  {author} {\bibinfo {author} {\bibfnamefont {A.}~\bibnamefont
  {Manchon}}, \bibinfo {author} {\bibfnamefont {H.~C.}\ \bibnamefont {Koo}},
  \bibinfo {author} {\bibfnamefont {J.}~\bibnamefont {Nitta}}, \bibinfo
  {author} {\bibfnamefont {S.~M.}\ \bibnamefont {Frolov}},\ and\ \bibinfo
  {author} {\bibfnamefont {R.~A.}\ \bibnamefont {Duine}},\ }\bibfield  {title}
  {\bibinfo {title} {New perspectives for {{Rashba}} spin--orbit coupling},\
  }\href {https://doi.org/10.1038/nmat4360} {\bibfield  {journal} {\bibinfo
  {journal} {Nature Materials}\ }\textbf {\bibinfo {volume} {14}},\ \bibinfo
  {pages} {871} (\bibinfo {year} {2015})}\BibitemShut {NoStop}%
\bibitem [{\citenamefont {von Oppen}\ \emph {et~al.}(2017)\citenamefont {von
  Oppen}, \citenamefont {Peng},\ and\ \citenamefont
  {Pientka}}]{oppenTopologicalSuperconductingPhases2017}%
  \BibitemOpen
  \bibfield  {author} {\bibinfo {author} {\bibfnamefont {F.}~\bibnamefont {von
  Oppen}}, \bibinfo {author} {\bibfnamefont {Y.}~\bibnamefont {Peng}},\ and\
  \bibinfo {author} {\bibfnamefont {F.}~\bibnamefont {Pientka}},\ }\bibfield
  {title} {\bibinfo {title} {Topological superconducting phases in one
  dimension},\ }in\ \href
  {https://doi.org/10.1093/acprof:oso/9780198785781.003.0009} {\emph {\bibinfo
  {booktitle} {Topological {{Aspects}} of {{Condensed Matter Physics}}:
  {{Lecture Notes}} of the {{Les Houches Summer School}}: {{Volume}} 103,
  {{August}} 2014}}},\ \bibinfo {editor} {edited by\ \bibinfo {editor}
  {\bibfnamefont {C.}~\bibnamefont {Chamon}}, \bibinfo {editor} {\bibfnamefont
  {M.~O.}\ \bibnamefont {Goerbig}}, \bibinfo {editor} {\bibfnamefont
  {R.}~\bibnamefont {Moessner}},\ and\ \bibinfo {editor} {\bibfnamefont
  {L.~F.}\ \bibnamefont {Cugliandolo}}}\ (\bibinfo  {publisher} {{Oxford
  University Press}},\ \bibinfo {year} {2017})\ p.~\bibinfo {pages}
  {0}\BibitemShut {NoStop}%
\bibitem [{\citenamefont {Buzdin}(2008)}]{buzdinDirectCouplingMagnetism2008}%
  \BibitemOpen
  \bibfield  {author} {\bibinfo {author} {\bibfnamefont {A.}~\bibnamefont
  {Buzdin}},\ }\bibfield  {title} {\bibinfo {title} {Direct {{Coupling Between
  Magnetism}} and {{Superconducting Current}} in the {{Josephson}}
  ${\ensuremath{\varphi}}_{0}$ {{Junction}}},\ }\href
  {https://doi.org/10.1103/PhysRevLett.101.107005} {\bibfield  {journal}
  {\bibinfo  {journal} {Physical Review Letters}\ }\textbf {\bibinfo {volume}
  {101}},\ \bibinfo {pages} {107005} (\bibinfo {year} {2008})}\BibitemShut
  {NoStop}%
\bibitem [{\citenamefont {Jauho}\ \emph {et~al.}(1994)\citenamefont {Jauho},
  \citenamefont {Wingreen},\ and\ \citenamefont
  {Meir}}]{jauhoTimedependentTransportInteracting1994}%
  \BibitemOpen
  \bibfield  {author} {\bibinfo {author} {\bibfnamefont {A.-P.}\ \bibnamefont
  {Jauho}}, \bibinfo {author} {\bibfnamefont {N.~S.}\ \bibnamefont
  {Wingreen}},\ and\ \bibinfo {author} {\bibfnamefont {Y.}~\bibnamefont
  {Meir}},\ }\bibfield  {title} {\bibinfo {title} {Time-dependent transport in
  interacting and noninteracting resonant-tunneling systems},\ }\href
  {https://doi.org/10.1103/PhysRevB.50.5528} {\bibfield  {journal} {\bibinfo
  {journal} {Physical Review B}\ }\textbf {\bibinfo {volume} {50}},\ \bibinfo
  {pages} {5528} (\bibinfo {year} {1994})}\BibitemShut {NoStop}%
\bibitem [{\citenamefont {{Mart{\'i}n-Rodero}}\ \emph
  {et~al.}(1994)\citenamefont {{Mart{\'i}n-Rodero}}, \citenamefont
  {{Garc{\'i}a-Vidal}},\ and\ \citenamefont
  {Levy~Yeyati}}]{martin-roderoMicroscopicTheoryJosephson1994}%
  \BibitemOpen
  \bibfield  {author} {\bibinfo {author} {\bibfnamefont {A.}~\bibnamefont
  {{Mart{\'i}n-Rodero}}}, \bibinfo {author} {\bibfnamefont {F.~J.}\
  \bibnamefont {{Garc{\'i}a-Vidal}}},\ and\ \bibinfo {author} {\bibfnamefont
  {A.}~\bibnamefont {Levy~Yeyati}},\ }\bibfield  {title} {\bibinfo {title}
  {Microscopic theory of {{Josephson}} mesoscopic constrictions},\ }\href
  {https://doi.org/10.1103/PhysRevLett.72.554} {\bibfield  {journal} {\bibinfo
  {journal} {Physical Review Letters}\ }\textbf {\bibinfo {volume} {72}},\
  \bibinfo {pages} {554} (\bibinfo {year} {1994})}\BibitemShut {NoStop}%
\bibitem [{\citenamefont {Liu}\ and\ \citenamefont
  {Chan}(2010)}]{liuAnomalousJosephsonCurrent2010}%
  \BibitemOpen
  \bibfield  {author} {\bibinfo {author} {\bibfnamefont {J.-F.}\ \bibnamefont
  {Liu}}\ and\ \bibinfo {author} {\bibfnamefont {K.~S.}\ \bibnamefont {Chan}},\
  }\bibfield  {title} {\bibinfo {title} {Anomalous {{Josephson}} current
  through a ferromagnetic trilayer junction},\ }\href
  {https://doi.org/10.1103/PhysRevB.82.184533} {\bibfield  {journal} {\bibinfo
  {journal} {Physical Review B}\ }\textbf {\bibinfo {volume} {82}},\ \bibinfo
  {pages} {184533} (\bibinfo {year} {2010})}\BibitemShut {NoStop}%
\bibitem [{\citenamefont {Sriram}\ \emph {et~al.}(2019)\citenamefont {Sriram},
  \citenamefont {Kalantre}, \citenamefont {Gharavi}, \citenamefont {Baugh},\
  and\ \citenamefont
  {Muralidharan}}]{sriramSupercurrentInterferenceSemiconductor2019}%
  \BibitemOpen
  \bibfield  {author} {\bibinfo {author} {\bibfnamefont {P.}~\bibnamefont
  {Sriram}}, \bibinfo {author} {\bibfnamefont {S.~S.}\ \bibnamefont
  {Kalantre}}, \bibinfo {author} {\bibfnamefont {K.}~\bibnamefont {Gharavi}},
  \bibinfo {author} {\bibfnamefont {J.}~\bibnamefont {Baugh}},\ and\ \bibinfo
  {author} {\bibfnamefont {B.}~\bibnamefont {Muralidharan}},\ }\bibfield
  {title} {\bibinfo {title} {Supercurrent interference in semiconductor
  nanowire {{Josephson}} junctions},\ }\href
  {https://doi.org/10.1103/PhysRevB.100.155431} {\bibfield  {journal} {\bibinfo
   {journal} {Physical Review B}\ }\textbf {\bibinfo {volume} {100}},\ \bibinfo
  {pages} {155431} (\bibinfo {year} {2019})}\BibitemShut {NoStop}%
\bibitem [{\citenamefont {Zeng}\ \emph {et~al.}(2003)\citenamefont {Zeng},
  \citenamefont {Li},\ and\ \citenamefont
  {Claro}}]{zengElectronicTransportHybrid2003}%
  \BibitemOpen
  \bibfield  {author} {\bibinfo {author} {\bibfnamefont {Z.~Y.}\ \bibnamefont
  {Zeng}}, \bibinfo {author} {\bibfnamefont {B.}~\bibnamefont {Li}},\ and\
  \bibinfo {author} {\bibfnamefont {F.}~\bibnamefont {Claro}},\ }\bibfield
  {title} {\bibinfo {title} {Electronic transport in hybrid mesoscopic
  structures: {{A}} nonequilibrium {{Green}} function approach},\ }\href
  {https://doi.org/10.1103/PhysRevB.68.115319} {\bibfield  {journal} {\bibinfo
  {journal} {Physical Review B}\ }\textbf {\bibinfo {volume} {68}},\ \bibinfo
  {pages} {115319} (\bibinfo {year} {2003})}\BibitemShut {NoStop}%
\bibitem [{\citenamefont {Sancho}\ \emph {et~al.}(1984)\citenamefont {Sancho},
  \citenamefont {Sancho},\ and\ \citenamefont
  {Rubio}}]{sanchoQuickIterativeScheme1984}%
  \BibitemOpen
  \bibfield  {author} {\bibinfo {author} {\bibfnamefont {M.~P.~L.}\
  \bibnamefont {Sancho}}, \bibinfo {author} {\bibfnamefont {J.~M.~L.}\
  \bibnamefont {Sancho}},\ and\ \bibinfo {author} {\bibfnamefont
  {J.}~\bibnamefont {Rubio}},\ }\bibfield  {title} {\bibinfo {title} {Quick
  iterative scheme for the calculation of transfer matrices: Application to
  {{Mo}} (100)},\ }\href {https://doi.org/10.1088/0305-4608/14/5/016}
  {\bibfield  {journal} {\bibinfo  {journal} {Journal of Physics F: Metal
  Physics}\ }\textbf {\bibinfo {volume} {14}},\ \bibinfo {pages} {1205}
  (\bibinfo {year} {1984})}\BibitemShut {NoStop}%
\bibitem [{\citenamefont {Wickramasinghe}\ \emph {et~al.}(2018)\citenamefont
  {Wickramasinghe}, \citenamefont {Mayer}, \citenamefont {Yuan}, \citenamefont
  {Nguyen}, \citenamefont {Jiao}, \citenamefont {Manucharyan},\ and\
  \citenamefont {Shabani}}]{wickramasingheTransportPropertiesSurface2018}%
  \BibitemOpen
  \bibfield  {author} {\bibinfo {author} {\bibfnamefont {K.~S.}\ \bibnamefont
  {Wickramasinghe}}, \bibinfo {author} {\bibfnamefont {W.}~\bibnamefont
  {Mayer}}, \bibinfo {author} {\bibfnamefont {J.}~\bibnamefont {Yuan}},
  \bibinfo {author} {\bibfnamefont {T.}~\bibnamefont {Nguyen}}, \bibinfo
  {author} {\bibfnamefont {L.}~\bibnamefont {Jiao}}, \bibinfo {author}
  {\bibfnamefont {V.}~\bibnamefont {Manucharyan}},\ and\ \bibinfo {author}
  {\bibfnamefont {J.}~\bibnamefont {Shabani}},\ }\bibfield  {title} {\bibinfo
  {title} {Transport properties of near surface {{InAs}} two-dimensional
  heterostructures},\ }\href {https://doi.org/10.1063/1.5050413} {\bibfield
  {journal} {\bibinfo  {journal} {Applied Physics Letters}\ }\textbf {\bibinfo
  {volume} {113}},\ \bibinfo {pages} {262104} (\bibinfo {year}
  {2018})}\BibitemShut {NoStop}%
\end{thebibliography}%

\end{document}